\documentclass[journal, comsoc]{IEEEtran}
\IEEEoverridecommandlockouts
\usepackage[utf8]{inputenc}
\IEEEoverridecommandlockouts
\usepackage[T1]{fontenc}
\usepackage[noend,lines numbered,ruled,vlined]{algorithm2e}
\usepackage{algcompatible}
\usepackage{graphicx}
\usepackage{textcomp}
\usepackage{multirow}
\usepackage{adjustbox}
\usepackage{array}
\usepackage{algpseudocode}
\usepackage{amsmath,amssymb,amsfonts}
\usepackage{float}
\usepackage[square, numbers, comma, sort&compress]{natbib}
\usepackage{mdframed}
\usepackage{amsmath}
\renewcommand{\arraystretch}{1.5}
\makeatletter
\newcommand\fs@norules{\def\@fs@cfont{\bfseries}\let\@fs@capt\floatc@ruled
  \def\@fs@pre{}%
  \def\@fs@post{}%
  \def\@fs@mid{\kern3pt}%
  \let\@fs@iftopcapt\iftrue}
\makeatother
\floatstyle{norules}

\usepackage{graphicx,caption}
\usepackage{ifmtarg}
\usepackage{subcaption}
\usepackage{enumitem}
\usepackage{diagbox}
\usepackage{makecell}

\usepackage{booktabs}
\usepackage{tikz}
\usetikzlibrary{graphs}
\usetikzlibrary{arrows.meta}
\usetikzlibrary{patterns}
\usetikzlibrary{positioning}
\usepackage{pgfplots}
    \pgfplotsset{
        table/search path={./plots},
    }
\usepackage{pgfplotstable}
\usepackage{wrapfig}
\usepackage[hyphens]{url}
\usepackage[hidelinks]{hyperref}
\hypersetup{breaklinks=true}
\usepackage{comment}
\usepackage{xcolor}
\def\BibTeX{{\rm B\kern-.05em{\sc i\kern-.025em b}\kern-.08em
    T\kern-.1667em\lower.7ex\hbox{E}\kern-.125emX}}
\usepackage{tabularx}
\usepackage{array}
\newcolumntype{P}[1]{>{\centering\arraybackslash}p{#1}}
\newcolumntype{M}[1]{>{\centering\arraybackslash}m{#1}}
\usepackage{url}
\usepackage{subcaption}
\usepackage{float}
\usepackage{stackengine}
\usepackage{mathtools}
\usepackage{longtable}
\usepackage{multicol}
\renewcommand{\arraystretch}{1.4}

\title{Blockchain Meets Adaptive Honeypots: A Trust-Aware Approach to Next-Gen IoT Security}

\author{
    \IEEEauthorblockN{
    Yazan Otoum\IEEEauthorrefmark{1}, Arghavan Asad\IEEEauthorrefmark{1}, Amiya Nayak\IEEEauthorrefmark{2} \\} \vspace{1em}
  \IEEEauthorblockA{\small \IEEEauthorrefmark{1}School of Computer Science and Technology, Algoma University, Canada}
        \\ \IEEEauthorblockA{\small \IEEEauthorrefmark{2}School of Electrical Engineering and Computer Science, University  of Ottawa, Canada}
    
        }
\date{February 2025}

\begin{document}

\maketitle

\footnotetext[1]{© 2025 IEEE. This work has been submitted to the IEEE Transactions on Network Science and Engineering for possible publication. Copyright may be transferred without notice, after which this version may no longer be accessible.}  

\begin{abstract}

Edge computing-based Next-Generation Wireless Networks (NGWN)-IoT offer enhanced bandwidth capacity for large-scale service provisioning but remain vulnerable to evolving cyber threats. Existing intrusion detection and prevention methods provide limited security as adversaries continually adapt their attack strategies. We propose a dynamic attack detection and prevention approach to address this challenge. First, blockchain-based authentication uses the Deoxys Authentication Algorithm (DAA) to verify IoT device legitimacy before data transmission. Next, a bi-stage intrusion detection system is introduced: the first stage uses signature-based detection via an Improved Random Forest (IRF) algorithm. In contrast, the second stage applies feature-based anomaly detection using a Diffusion Convolution Recurrent Neural Network (DCRNN). To ensure Quality of Service (QoS) and maintain Service Level Agreements (SLA), trust-aware service migration is performed using Heap-Based Optimization (HBO). Additionally, on-demand virtual High-Interaction honeypots deceive attackers and extract attack patterns, which are securely stored using the Bimodal Lattice Signature Scheme (BLISS) to enhance signature-based Intrusion Detection Systems (IDS). The proposed framework is implemented in the NS3 simulation environment and evaluated against existing methods across multiple performance metrics, including accuracy, attack detection rate, false negative rate, precision, recall, ROC curve, memory usage, CPU usage, and execution time. Experimental results demonstrate that the framework significantly outperforms existing approaches, reinforcing the security of NGWN-enabled IoT ecosystems.\\

\textbf{Keywords—} Blockchain, Adaptive Honeypots, IoT Security, Intrusion Prevention, Trust-Aware Networks, Next Generation IDS.
\end{abstract}

\section{Introduction}

The rapid proliferation of IoT applications has been facilitated by advancements in NGWN. These networks enable efficient and large-scale service provisioning through technologies such as Mobile Edge Computing (MEC), which enhances computational efficiency by processing data closer to users \cite{zhang2020real}. However, the increasing connectivity and integration of IoT devices within NGWN have exposed these networks to a wide range of cybersecurity threats \cite{wazid2020security}. Traditional intrusion detection and prevention mechanisms struggle to provide robust security due to the constantly evolving attack strategies employed by adversaries \cite{nilkanthsing2020dynamic}. A recent 2024 cyberattack on a 5G-enabled IoT network disrupted critical healthcare and industrial services, underscoring the urgency of strengthening security in NGWN-based IoT environments. IDS play a crucial role in securing IoT networks, with detection methods categorized as Signature-based IDS (SIDS) and Anomaly-based IDS (AIDS). SIDS typically provide high detection accuracy by comparing incoming network traffic against predefined attack signatures. However, they fail to detect novel and unknown attacks \cite{aravamudhan2021survey}. On the other hand, AIDS attempts to identify attacks by extracting and analyzing various traffic features, but its effectiveness is often hampered by limited generalization ability and computational inefficiency \cite{chaabouni2019network}. While deep learning-based anomaly detection has been explored to improve detection accuracy \cite{al2025comprehensive}, classification based solely on network flow features has proven to be insufficient \cite{hinojosa2024edge}. Moreover, existing IDS approaches do not adequately address the need for real-time detection with minimal false positives and high adaptability to evolving attack strategies \cite{toony2024multi, saheed2024modified}. Blockchain technology has emerged as a promising solution to enhance security in IoT environments by ensuring tamper-resistant and decentralized trust management \cite{maurya2025blockchain}. While blockchain-based IDS solutions mitigate various security threats, they introduce concerns regarding computational overhead, transaction latency, and scalability, particularly in resource-constrained environments \cite{otoum2024advancing}, \cite{dawit2020suitability}. Furthermore, deploying multiple network entities for monitoring and intrusion detection increases the complexity of security management \cite{louati2020deep}. The deployment of honeypots as deception mechanisms has proven effective in attracting attackers and collecting attack patterns to improve network security \cite{franco2021survey}. High-interaction honeypots engage attackers longer, allowing for comprehensive analysis of attack strategies \cite{khan2020reputation}. However, existing honeypot-based approaches either deploy static honeypots or fail to dynamically adapt to emerging threats. Additionally, game-theoretic models proposed for attack detection \cite{zheng2025predictive} often assume perfect knowledge of attacker strategies, which is unrealistic in real-world scenarios. Despite advancements in IDS and security frameworks, critical security challenges persist in NGWN-enabled IoT networks due to the following limitations:

\begin{itemize}
\item High Latency: Traditional IDS approaches rely on centralized cloud processing, introducing delays in attack detection and response.
\item High False Alarm Rate: Anomaly detection models often produce a high number of false positives due to insufficient training on diverse attack scenarios.
\item Lack of Adaptive Defense Mechanisms: Existing solutions lack dynamic attack mitigation strategies, allowing attackers to bypass static security measures.
\end{itemize}

To address these challenges, we propose a trust-aware security framework that integrates blockchain authentication, advanced intrusion detection, and dynamic honeypot deception. Our approach is designed to enhance real-time threat mitigation while minimizing computational overhead. The key contributions of this research include:
\begin{itemize}
\item A blockchain-based authentication mechanism that ensures the legitimacy of IoT devices and prevents impersonation attacks using the DAA.
\item A bi-stage intrusion detection framework that combines IRF-based SIDS with DCRNN-based AIDS, improving detection accuracy while maintaining efficiency.
\item A moving target defence strategy that dynamically migrates services to trusted edge nodes based on trust evaluation, mitigating the risk of service compromise.
\item The deployment of high-interaction, on-demand virtual honeypots to deceive attackers, extract attack signatures, and enhance IDS training using securely stored attack patterns.
\end{itemize}

The proposed approach is implemented and evaluated in the NS3 simulation environment, demonstrating a 25\% improvement in detection accuracy, a 30\% reduction in false negatives, and enhanced resource efficiency compared to existing methods. The remainder of this paper is organized as follows: Section II gives the background. Section III provides a comprehensive literature review of existing intrusion detection and prevention mechanisms. Section IV outlines the problem statement and research challenges. Section V presents the proposed methodology in detail. Section VI discusses the experimental setup and performance evaluation. Finally, Section VII concludes the paper and outlines directions for future work.

\section{Background}
This section discusses the background of Blockchain technology, the Need for honeypot deployment, and integrated blockchain honeypot deployment to improve security attack mitigation in an IoT environment \cite{otoum2024enhancing}. Blockchain is an open-access ledger of digital transactions for public networks where the attacker cannot change or compromise. It is a decentralized platform managed by several authorities and mitigates the issue of less security caused by the centralized model. It allows its users to deal directly with each other and reduces security issues caused by third parties. Blockchain technology in the field of IoT demands high levels of demand because of its decentralized nature. All transactions, such as authentication, network access, etc., are stored in blockchain because of its resource-constraint nature and tamperproof nature, which allows its users to verify by digital transaction. Information collected by the users is chained by blocks using a ledger, which is tightly chained with each other with the help of hashed cryptographic keys. The most common method to store transactions in the blockchain is the Merkle tree, which stores the transactions done by individuals, and the root hash tree, which stores them in the blockchain. Each transaction is hashed in a cryptographic manner. Miners are used to store the upcoming transactions, which generate a key for every user and allow them to join the ledger. The data in the blockchain cannot be modified or hacked, providing that it has an immutable character. Some of the other significant features of blockchain are fault tolerance, transparency, improved security, and decentralized nature. 

Honeypot is an intelligent attack detection technology that shows the attacker's intention and reports the system's status, whether there was an abnormal behaviour or an external attacker in the network environment and captures the way they attack. It is the copied environment of the original environment with the same configurations that attracts the attackers to hack and record their patterns to store them in the log files. The storage of attack patterns helps to drop the malicious users in the future, who will be identified as attackers based on their patterns, which improves the overall safety of the system.

A secure and reliable IoT system must integrate blockchain and honeypot technology to improve IDS security. Physically deployed honeypots are used for IDS, which is effective and improves the overall system safety; however, the attackers may compromise them, or the attackers might have awareness about the honeypot, which creates risk in the IoT environment. By using physically deployed honeypots, the overall cost of the system would be high, leading to more energy consumption. To overcome the above challenges, virtual honeypots are deployed at the time of attack, which improves overall safety and low cost. Table~\ref{FEATURES OF BLOCKCHAIN AND HONEYPOT} provides the distinct characteristics of blockchain and honeypot technologies. Hence, in our approach, we integrate the blockchain honeypot methodology in IDS to detect and prevent intrusions in the NGWN-IoT environment. 

\begin{table*}[ht]
\caption{FEATURES OF BLOCKCHAIN AND HONEYPOT}
\centering
\renewcommand{\arraystretch}{1.2}
\begin{tabular}{|p{1.2in}|p{1.9in}|p{1.9in}|}
\hline
\textbf{Features} & \textbf{Blockchain} & \textbf{Honeypot} \\
\hline
Security & 
\parbox[t]{1.9in}{Provides high security due to its immutable feature. Privacy is protected using hashing.} & 
\parbox[t]{1.9in}{Security is provided by attracting attackers and learning their strategies.} \\ \hline

Risk & 
\parbox[t]{1.9in}{Low risk due to robustness.} & 
\parbox[t]{1.9in}{Moderate risk when relying solely on a honeypot for security.} \\ \hline

Information Gathering & 
\parbox[t]{1.9in}{Information is stored in blocks after consensus validation.} & 
\parbox[t]{1.9in}{Information gathered is stored in log files, revealing attacker strategies.} \\ \hline

If Integrated with IDS & 
\parbox[t]{1.9in}{Provides authenticity and security.} & 
\parbox[t]{1.9in}{Helps in intrusion prevention by learning new attack strategies.} \\ \hline

\end{tabular}
\label{FEATURES OF BLOCKCHAIN AND HONEYPOT}
\end{table*}

\section{LITERATURE SURVEY}

This section surveys the literature on several intrusion detection and prevention schemes, including Artificial intelligence (AI)- based schemes and honeypot-based schemes. It analyzes the objectives, workings, and research gaps of these approaches.

\subsection{Intrusion Detection and Prevention Schemes}

An intrusion detection and prevention system for NC-assisted mobile small cells was proposed in \cite{parsamehr2019novel}. Initially, the nodes divide the message into a sequence of packets. The proposed work has four steps; the first is the tag generation process that calculates the tag value for every coded packet. The second one is the swapping process, which avoids tag pollution attacks from the coded packets, in which the source node swaps the tag and coded packets.  The third one is the key distribution process; here, the maximum counts of key vectors are assigned for every intermediate and destination node. Finally, the verification process verifies the swapped coded packets. During verification, if the result equals zero, the packet will be sent to the next level; otherwise, it will be dropped. However, this approach uses a single security parameter for key generation, which is insufficient for security; hence, it will be easily compromised by the attackers, leading to poor security. Multi-agent assisted IDS in IoT environment using Blockchain technology was proposed in \cite{liang2020intrusion}. Three operations were used in this paper: data collection, data management, and data response. For testing the proposed work, the NSL-KDD dataset is used to detect attacks from the transport layer. For attack detection reasons, the NSL-KDD dataset is used, and the agents used for intrusion detection are as follows: communication agent, response agent, collection agent, host agent, detection agent, and training agent. Based on the attack information, the database is trained for the network. However, the deployment of a larger number of agents increased the network complexity. The computational overhead for running all these agents over the blockchain increases transaction verification and hash generation times.

A practical approach for automatically protecting NGWN mobile networks using multiple tenants was proposed in \cite{mamolar2019autonomic}. The network architecture is comprised of a radio access segment, edge computing segment, core network segment, and interdomain segment. The dynamic traffic of the NGWN network was considered, and an effective IDS was proposed. The network is divided into two types, namely, data network and management network. The automatic detection of intrusions is executed using several agents, namely flow control agents, action enforcers, decision-makers, and security monitoring agents. The packets in the traffic flow were classified by the security monitoring agent, which was responsible for the generation of alerts for the decision maker to decide the corresponding action. The significant attacks mitigated by this approach are UDP flooding attacks and DDoS attacks.

A multi-view consistent Generative Adversarial Network (GAN) model was designed to enhance Intrusion Detection and Prevention Systems (IDPS) for IoT \cite{rajkumar2025multi}. This model leverages multiple data views to improve the detection of security threats. A practical framework for detecting intrusions in the IoT based on real-time dataset was proposed in \cite{al2020real}. The possibility of attacks on several aspects of the IoT network was described, and the necessity of an effective intrusion detection system was demonstrated. The effect of the dataset's quality, which affects the performance of IDS, was evaluated to design a real-time dataset. The proposed dataset consisted of data obtained from various IoT nodes containing attack patterns of known attacks. Some of the attacks considered in this approach were flooding, black holes, and selective forwarding attacks. Thereby, the users were able to obtain the dataset to perform efficient intrusion detection. Several known attacks were identified by utilizing the real-time dataset, but the features extracted for identification were not sufficient to perform the detection of other complex intrusions.

An intrusion mitigation technique for mitigating attacks in the IoT environment was proposed in \cite{hatzivasilis2019wardog}. The authentication of the user is carried out to allow only legitimate users to participate in data transmission. Further, an ongoing traffic analysis was carried out based on the packet flow features. This work undergoes three stages in which the log files of the authenticated entity are examined to detect attacks caused by bots. Then, the attacker was detected by analyzing the typical pattern of attack, thereby isolating the attacker from the network. An encryption-based approach for secure data collection, storage and access in cloud-assisted IoT was proposed in \cite{wang2018secure}. The issues in the proper management of data in cloud storage were addressed. The IoT sensors were used for periodic data acquisition from the environment and stored in cloud storage. Here, the attacker in the network was able to compromise data to exploit its integrity. For this purpose, the authors proposed a conditional identity-based broadcast proxy re-encryption (CIBRE) to maintain the confidentiality of the data. The re-encryption of data was also provided to ensure the integrity of the cloud data each time it was decrypted. Identity-based encryption was implemented to ensure the confidentiality of the data, but entities such as cloud storage can be compromised by malicious attackers, causing a single point of failure.

\subsection{AI-based intrusion detection schemes}

\begin{table*}[!p] 
\scriptsize
\centering
\caption{RESEARCH GAPS IN LITERATURE SURVEY}
\renewcommand{\arraystretch}{1.9} 

\begin{tabular}{|p{1.7cm}|p{2.4cm}|p{4cm}|p{4.5cm}|p{4.2cm}|}
\hline
\textbf{Schemes} & \textbf{Reference} & \textbf{Objective} & \textbf{Algorithms/Methodology} & \textbf{Disadvantages} \\ 
\hline
\multirow{7}{=}{\centering Intrusion Detection and Prevention Schemes} 
& Parsamehr, R. et al. \cite{parsamehr2019novel} & Intrusion detection and prevention system for NC-assisted mobile small cells 
& Four-step approach (Tag generation, Swapping, Key distribution, and Verification) 
& 1. Poor Security \newline 2. Less efficiency in detecting malicious users \\ \cline{2-5}
& Liang, C. et al. \cite{liang2020intrusion}
& Multi-agent assisted IDS in IoT environment using Blockchain technology 
& Agents-based approach (Communication agent, Response agent, Collection agent, Host agent, Detection agent, and Training agent) 
& 1. High Complexity \newline 2. High Computational overhead \newline 3. Lack of Blockchain authentication parameters \\ \cline{2-5}

& Mamolar, A. S. et al. \cite{mamolar2019autonomic}
& Automatic protection of 5G mobile networks using multiple-tenants 
& Agents-based approach (Flow control agents, Action enforcer, Decision maker, and Security monitoring agent) 
& 1. Energy consumption due to multiple agents \newline 2. Less secure \\ \cline{2-5}
& Al-Hadhrami, Y. et al. \cite{al2020real}
& Framework for the detection of intrusions in the IoT based on real-time dataset 
& Dataset-based approach (Flooding attack, Black hole attack, and Selective forwarding attack) 
& 1. Lack of features \newline 2. High Computation required \\ \cline{2-5}
& Hatzivasilis, G. et al. \cite{hatzivasilis2019wardog}
&  Intrusion mitigation technique for mitigation of attacks in the IoT environment
& Three-stage approach
(Authentication, Detection and Security) 
& 1. Inefficient legitimate users’ authentication \newline 2. Inefficiency in malicious user classification and detection \\ \cline{2-5}
& Wang, W. et al. \cite{wang2018secure}
& Encryption-based approach for secure collection of data, storage and access in cloud-assisted IoTs
& Conditional Identity-based Broadcast proxy RE-encryption (CIBRE) method
& 1. Single point failure \newline 2. No transitivity by using CIBRE \newline 3. Poor security \\ \cline{2-5}
& Rajkumar, M. et al. \cite{rajkumar2025multi}
&  Generative adversarial networks (GANs) based approach to enhance protection against security threats in IoTs
& Developing a GAN architecture to learn from multiple data gathered from various sources in IoT simultaneously
& 1. High Computational Complexity \newline 2. Challenges in obtaining diverse and qualified data dynamically \newline 3. Poor Scalability \\ \hline
\multirow{8}{=}{\centering AI-based Intrusion Detection Schemes} 
& Maimó, L. F. et al. \cite{maimo2018self} 
& Deep learning method for detecting network anomalies by extracting features from the network traffic in 5G network 
& Four-step approach
(Virtual infrastructure, Virtualized network functions, Management and Orchestration)
& 1. High latency \newline 2. High complexity \newline 3. Low classification accuracy \\ \cline{2-5}
& Lam, J. et al. \cite{lam2020machine} 
& Machine learning algorithm for detecting network anomalies in 5G network 
& Machine Learning Algorithm
& 1. Time complexity due to lots of training required \newline 2. Lack of classification parameters \\ \cline{2-5}
& Yang, A. et al. \cite{yang2019design} 
& Intrusion detection system for IoT environment using improved BP neural network 
& LM-BP Algorithm
& 1. Attack detection inefficiency \newline 2. Sensitive to noise \newline 3. Poor security \\ \cline{2-5}
& Mondal, A. et al. \cite{mondal2021enhanced} 
& Security approach for cloud storage based on encryption mechanism
& Co-occurrence Matrix Algorithm and cryptographic algorithm 
& 1. Required several training data\newline 2. Inefficient honeypot adoption leads to less security \\ \cline{2-5}
& Eskandari, M. et al. \cite{eskandari2020passban} 
& Gateway-based detection of network intrusions 
& Gateway-based (Timings involved in the traffic flow and Data flow statistics over a particular time interval)
& 1. Pour scalability \newline 2. Inefficient features classification for intrusion detection leads to several attacks \\ \cline{2-5}
& Li, B. et al. \cite{li2020deepfed}
& Intrusion detection scheme in industrial cyberphysical systems
& Three model approach
 (Trusted authority, Industrial agents and Cloud server)
& 1. High latency \newline 2. High computation due to several agents\\ \cline{2-5}
& Mahdi, M. A. et al. \cite{mahdi2024secure} 
& Implementing a trained hybrid machine learning technique based on the extracted features from live IoT traffic 
& Hybrid (Decision trees-support vector machines-neural networks)
& 1. High Computational Overhead \newline 2. Gathering Large amounts of high-quality data from various IoT devices to train effectively \\ \cline{2-5}
& Mallidi, S. K. R. et al. \cite{mallidi2025advancements} 
& Examining different machine learning models and integrating them into IoT systems for effective intrusion detection
& Systematically analyze and summarize existing research on the training and deployment strategies of AI-based IDS in IoT
& 1. Complexity \newline 2. Data privacy concern \newline 3. Frequent retraining and updates \\ \hline
\multirow{5}{=}{\centering Honeypot-Based Intrusion Detection and Prevention Schemes} 
& Al-Mohannadi, H. et al. \cite{al2020analysis}
& Honeypot-based analysis of cybersecurity attacks 
& Parameters-based approach (IP address, Domain name, Username, Password, and Geographic location) 
& 1. Lack of authentication parameters leads to malicious user registration \newline 2. Honeypot creation seems to be less secure \\ \cline{2-5} 
& Lee, J. et al. \cite{lee2020phantomfs} & Detection of malicious users using a file-based deception technique 
& Three-step approach (Regular files, Fake files, and Sensitive files) 
& 1. Files are easily compromised if containing sensitive data \newline 2. High latency due to file generation during honeypot deployment \\ \cline{2-5} 
& Li, B. et al. \cite{li2020anti} & Efficient approach for learning attack strategies using a honeypot-based deception strategy 
& Hybrid Game Theoretic-based Approach (One player and ICPS defender) 
& 1. Leads to security threats \newline 2. Does not consider all types of attacks \\  \cline{2-5} 
& Dara, N. et al. \cite{dara2024intelligent} & Deployment of honeypots—decoy systems mimicking real IoT devices designed to attract potential attackers
& Establishing a honeypot designed for IoT devices to provide deeper insights into IoT threats
& 1. High resource intensive \newline 2. Limited Coverage \newline 3. Lack of Real-Time Detection \\ \cline{2-5} 
& Ntizikira, E. et al. \cite{ntizikira2024honey} & Honeypot technology with blockchain principles combination 
& Integration of edge computing with ensemble learning models
& 1. Poor Scalability \newline 2. Real-time Processing Constraints \newline 3. High Cost and Resource Overhead\\  \hline
\end{tabular}
\label{RESEARCH GAPS IN LITERATURE SURVEY}
\end{table*}

A deep learning method for detecting network anomalies by extracting features from the network traffic in NGWN networks was proposed in \cite{maimo2018self}. The proposed architecture has four components: virtual infrastructure, virtualized network functions, management and orchestration, and operations and business support systems. Initially, the network flow is collected from the user equipment. From the network flow, features are extracted by a feature vector. Based on the extracted features, the anomaly symptoms are detected and sent to the LSTM for recognizing the anomaly patterns.  The simulation result shows that the proposed system achieves high performance in terms of anomaly detection accuracy. This approach takes only flow-based features for anomaly detection, thus reducing classification accuracy. It takes all the packets as input, therefore increasing the latency and complexity of the process. A machine learning algorithm for detecting network anomalies in a 5G network was proposed in \cite{lam2020machine}. The data are pre-processed to enhance detection accuracy.  The features are extracted from the pre-processed data. Weibull distribution is used to select the features from the normal traffic.  The Convolutional Neural Network (CNN), which was designed with the NAS method, is used to classify anomalies using features. The proposed work achieves high accuracy and low latency by optimizing the neural network in the 5G network. A hybrid machine learning technique (decision trees-support vector machines-neural networks) to improve detection accuracy and robustness against various types of abnormal or malicious activities in IoT was proposed in \cite{mahdi2024secure}. Various training methodologies for AI-based IDS, proposed in \cite{mallidi2025advancements}, highlighted the importance of selecting appropriate datasets, feature selection techniques, and model training processes to enhance IoT accuracy and efficiency. The paper discussed deployment strategies, including integrating IDS into IoT networks, real-time monitoring capabilities, and the scalability of AI models to handle the dynamic nature of IoT environments. An intrusion detection system for an IoT environment using an improved BP neural network was proposed in \cite{yang2019design}. The proposed system used the LM-BP algorithm for intrusion detection. Initially, the data are collected from the data source. From the collected data, features are extracted using the proposed neural network. Here, the neural network was divided into two sections; one section was used to optimize the connection weight value, and another was used to optimize the learning rate, which adjusts dynamically to improve the detection rate. It was used to detect intrusion behaviour from the extracted features. After detecting the intrusions, the neural network was updated. The proposed system detected DoS, R2L, U2L and probing attacks using the KDD CUP 99 dataset. However, in this approach, only particular attacks were detected, and the prevention of these attacks was not investigated, which affects the overall security of the system. A practical security approach for cloud storage based on an encryption mechanism was proposed in \cite{mondal2021enhanced}. Initially, the data was normalized in the dataset to improve the accuracy by removing the unwanted data and processing the missing data. Then, the grey-level co-occurrence matrix algorithm was implemented to extract the significant features from which attacks were classified using a CNN-based classifier. The security of the data was ensured by adopting a honeypot-based cryptographic algorithm, which is responsible for performing encryption or decryption of data. The gateway-based detection of network intrusions was presented in \cite{eskandari2020passban}. Generally, it needs a scalable dataset which enables the addition of new attacks in various settings. Dynamic features were considered on the gateway to extract the features for normal and abnormal classification. It avoids considering features which were static concerning the environment; the features considered were related to the flow of packets, timings involved in the traffic flow and data flow statistics over a particular time interval. Since Passban was an anomaly-based IDS, it could be trained while observing the target system's network traffic in a normal state. 

An intrusion detection scheme in industrial cyber-physical systems based on federated deep learning was proposed in \cite{li2020deepfed}. The system model comprises three main entities: trusted authority, industrial agents and cloud servers. The industrial agents represent the industrial administrators responsible for improving the intrusion detection system. The trusted authority was responsible for key generation to authenticate both the cloud server and industrial agents. The threat model, consisting of several significant threats, such as command injection attacks, response injection attacks, reconnaissance attacks and DoS attacks, was explained. The deep-fed scheme was proposed, which comprised the integration of CNN and Gated Recurrent Unit (GRU) based intrusion detection system and secure communication protocol based on Pallier cryptosystem was executed. 

\subsection{Honeypot-based intrusion detection and prevention schemes}

The honeypot-based analysis of cyber security attacks was proposed in \cite{al2020analysis}. Cyber security threats are identified to detect and mitigate the behaviours of web services. Information Collected from the data included IP address, domain name, user name, password and geographic location of the attacker. Due to the centralized environment, it was complex to identify and attract attackers. For intrusion detection, honeypot technology is used, which mimics real-time systems and determines the attacks. Finally, a log file report is generated for the collected data and stored in the cloud servers. 

An approach for detecting malicious users using a file-based deception technique was proposed in \cite{lee2020phantomfs}. A hidden interface was established for legitimate users, which malicious users cannot access. The system files are divided into three types, namely, regular files, fake files and sensitive files, in which the regular files and fake files are accessible for both the regular interface and hidden interface. The sensitive files were accessible only to legitimate users who communicate through a hidden interface. The honeypot-based deception technique was implemented in which the malicious users were attracted to attack the honey files, which are referred to as fake files, from which the behaviour of attackers, including their resources, was identified to improve the security of the sensitive files. An efficient approach for learning the attack strategies using a honeypot-based deception strategy for industrial cyber-physical systems was proposed in \cite{li2020anti}. Here, the attack strategies of attackers after identifying the honeypot were studied to provide overall security to the ICPS.
The application of advanced analytics techniques to the data gathered from honeypots, aiming to identify and understand emerging threat patterns in IoT environments, was proposed in \cite{dara2024intelligent}. Here, a machine learning model is used to analyze attackers' behaviour captured by honeypots and identify subtle, previously unseen anomalies that may signal new or evolving attack strategies.
An edge-assisted ensemble learning model within the honey-block framework to proactively identify and mitigate potential security breaches for intrusion detection in IoT was proposed in \cite{ntizikira2024honey}. This framework combines the power of edge computing and ensemble learning to create a more secure environment for IoT systems. By aggregating predictions from various models, ensemble methods reduce the likelihood of false positives and enhance the system's ability to detect complex, previously unknown attacks.

A hybrid game theoretic-based approach was implemented, and the players had different objectives. The two-player game was proposed with attackers as one player and ICPS defenders as another player. The one-shot signalling model was developed. The perfect Bayesian equilibrium-based probability was calculated for the successful defence of all the attacks provided by the attackers. However, performing only honeypot-based intrusion detection in the ICPS is not enough to mitigate all types of attacks that threaten the system's security. Table~\ref{RESEARCH GAPS IN LITERATURE SURVEY} provides the working of existing approaches from which the research gaps in the literature survey are determined based on methodology and disadvantages.

\section{PROBLEM STATEMENT}

This section deals with the major problems encountered in the existing approaches to detecting and preventing cyber-attacks in the NGWN-IoT environment. These problems are considered as the problem statement of our approach, and the research solutions provided by our approach are also mentioned. The IoT devices' legitimacy was considered a significant property to be ensured to mitigate several cyber-attacks in the IoT environment \cite{jangirala2019designing}. The RFID-based authentication was implemented to ensure the security of transmitted data. The RFID credentials of each entity were registered, and a public address was generated. The lightweight authentication was performed by implementing simple bit-wise exclusive OR functions. The problems faced by this approach are mentioned as follows,

\begin{itemize}
    \item The authentication of entities was performed based on RFID tags, but considering only one credential for authentication limits the system's security as the attackers could manipulate the RFID.
    
    \item The RFID-based authentication doesn’t capture unknown attacks, thereby increasing the attackers' involvement rate to launch more attacks in the IoT user environment.
    
\end{itemize}

Several game theoretic models were proposed for intrusion detection and prevention in the cloud server. The risk level of virtual machines in the cloud server was computed, and the honeypot-based analysis of the attack pattern was performed \cite{wahab2019resource}. The game model for intrusion detection in the cloud was executed, which carried out various types of IDS such as SIDS, AIDS, and honeypot-based IDS \cite{gill2020gtm}. However, these approaches encounter several problems that are mentioned as follows:

\begin{itemize}
    \item Since all the processes take place in a cloud environment, there is a great threat to the security of the cloud, and the centralized nature of the cloud environment results in a single point of failure and violation of quality of service and SLA. 

    \item The game theory-based IDS model detected attacks by performing signature-based, anomaly-based and honeypot-based IDS, but the overall security of the cloud computing was not ensured as it can be compromised through malicious users.

    \item The intrusion detection was implemented as a game theory-based model in which if the number of attacks increased, the approach's complexity also increased, thereby increasing the latency.
    
\end{itemize}

The honeypot-based detection and mitigation of cyberattacks were performed using various approaches. The rule-based determination of malicious behaviour possessed by the data packet was presented in \cite{chakkaravarthy2020design}. The multiport honeynet model was designed with several interactive level honeypots to extract information about the attacker strategy \cite{zhang2019iot}. The problems faced by these approaches are mentioned as follows:

\begin{itemize}
    \item Only honeypot-based intrusion detection was carried out in which the attacker was assumed not to be aware of honeypots. But if the attacker identifies a honeypot and performs the attack in the system file by omitting it, it will be useless.
    
    \item The type of honeypot used in this approach was a low interaction honeypot, which will obtain only less information about the attacker pattern, which was not enough to perform effective intrusion detection. 

    \item The log files are generated on the corresponding servers to improve the security of the SOAP ports. However, these files can also be compromised by the attackers, affecting the integrity of log files.
\end{itemize}

The proposed approach is designed to mitigate all the research problems encountered in the existing approaches. In our approach, the blockchain-based authentication of IoT devices is performed based on attributes such as PUF, device ID and MAC address. These attributes are unchangeable and hence provide a high degree of security. The SIDS and AIDS models are utilized based on attack patterns and spatiotemporal features to detect known and unknown attacks. The deployment of global and local mobile edge computing for provisioning services was carried out, which mitigates the limitation of increased latency and performs in a decentralized manner. The service provisioning is executed in the mobile edge computing node, where the migration of services is computed based on trust calculation. As the process is carried out in the edge node, the time consumed is reduced, fulfilling the quality of service and SLA constraints. The on-demand placement of virtual high-interaction honeypots is carried out to attract the attackers and analyze their strategies. These strategies are stored in the log files in an encrypted manner and are used to train the SIDS model.

\section{PROPOSED MODEL}

\begin{figure*}
    \centering
    \includegraphics[width=0.7\textwidth, height=0.7\textwidth]{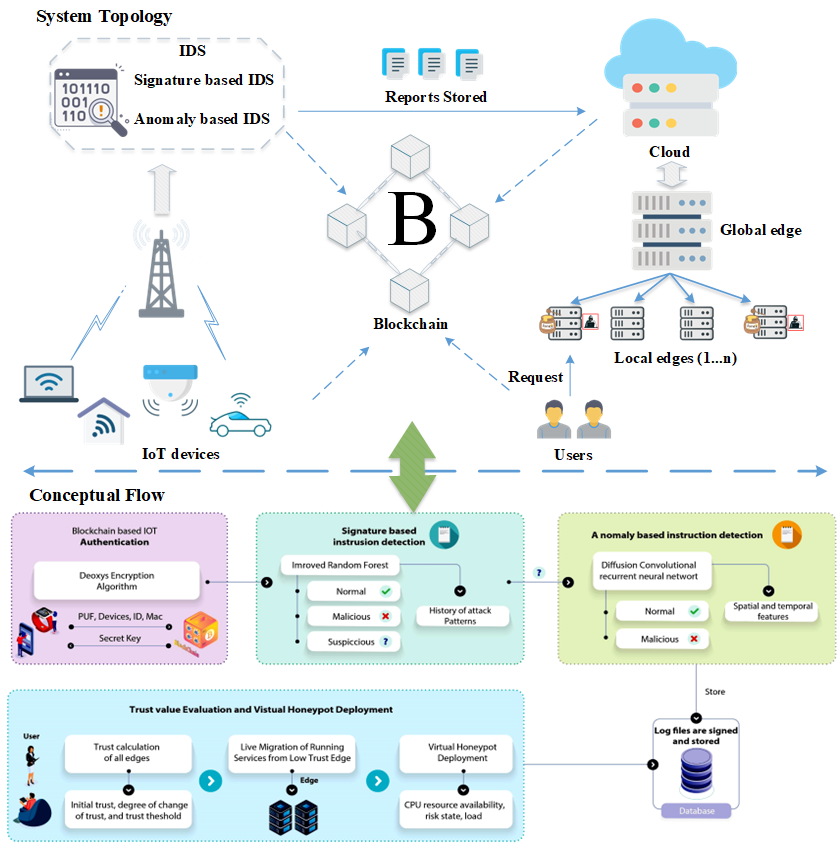}
    \caption{Model topology and proposed architecture}
    \label{fig1}
\end{figure*}

\begin{table}[ht]
\caption{TABLE OF NOTATIONS}
\centering
\renewcommand{\arraystretch}{1.2}
\begin{tabular}{|c|l|}
\hline
\textbf{Notation} & \textbf{Description} \\
\hline
$D_{ID}$ & Device ID \\ \hline
$U_{ID}$ & User ID \\ \hline
$K_{pr}$ & Secret Key \\ \hline
$K_{tg}$ & Tag \\ \hline
$K_L$ & Key lifetime \\ \hline
$\Theta_0$ & Initial forest \\ \hline
$Z_0$ & Number of Trees \\ \hline
$\varsigma_0$ & Feature vector \\ \hline
$W^T (k)$ & Weight of the feature K \\ \hline
$\alpha^T$ & Weight of the tree T \\ \hline
$\Delta$ & Important feature bags \\ \hline
$\delta h$, $\delta g$ & Variation of important and unimportant features \\ \hline
$M_{av}$ & Average number of nodes in a tree \\ \hline
$I_l$ & Input image length \\ \hline
$Q$ & Number of zeroes \\ \hline
$K_r$ & Size of the kernel \\ \hline
$SK$ & Kernel stride \\ \hline
$tanh$, $\lambda $ & Activation functions \\ \hline
$\tau(E)$ & Trust value of edge server \\ \hline
$In_\tau$ & Initial trust value \\ \hline
$\vartheta$ & Degree of change of trust \\ \hline
$RA_{CPU}$ & CPU resource availability \\ \hline
$E_i$, $T_i$ & Number of edge servers, Number of tasks \\ \hline
$SA_N^{DIM}$ & N search agent with dimensions \\ \hline
$FV$ & Fitness value \\ \hline
$R_n$, $B_n$ & Root node and branch nodes \\ \hline
$G_{2P}^{nxm}$ & Gaussian samples with matrix size and prime \\ \hline
$N_\sigma^n$ & Gaussian distribution with n dimensions \\ \hline
$K_c$ & Shift vector \\ \hline
$re$ & Random bit \\ \hline
$d_s$ & Sample vector \\ \hline
$F$ & Challenge vector \\ \hline
$\overrightarrow{R}$, $\overrightarrow{Bn}$ & Index vector position \\ \hline
$T(ft)$ & Temporary fitness value \\ \hline
\end{tabular}
\label{NOTATION TABLE}
\end{table}


In this work, we focus on detecting and preventing intrusions in an IoT environment. Here, security and privacy of the IoT environment are accomplished for both IoT devices and users by using honeypots and blockchain technologies. Fig.~\ref{fig1} illustrates the system topology and conceptual flow of the proposed approach, and all the notations and parameters used in this paper are shown in Table~\ref{NOTATION TABLE}. 

\subsection{System Model}

The system model comprises a set of IoT devices $D=\{D_1,D_2,...,D_n\}$ from which the data are generated and sent to the cloud server for processing and storage purposes. The IoT environment consists of a set of IoT users $U=\{U_1,U_2,...,U_n\}$ who request service from the cloud server. The proposed IDS is deployed between the IoT devices and the cloud server for filtering data packets from the devices through the 5G Base Station (BS). The edge layer comprises a global edge server (GE) and a set of local edge servers $LE=\{LE_1,LE_2,...,LE_n\}$ that are placed at the edge of the network to provide services to the IoT users with optimal QoS, quality of service SLA requirements. The attackers in the system are considered on both the IoT devices side and the user side, who try to affect the provisioning of legitimate services to the users. On the IoT devices side, the attackers deploy a malicious IoT device or compromise the legitimate IoT devices to generate malicious packets. On the other hand, the attackers on the user side try to compromise the user device or directly local edge servers to affect the provisioning of services. The processes involved in the proposed approach to perform secure service provisioning in the IoT environment are described as follows.

\subsection{Blockchain-based IoT authentication}

\begin{figure}
    \centering
    \includegraphics[height=.255\textwidth]{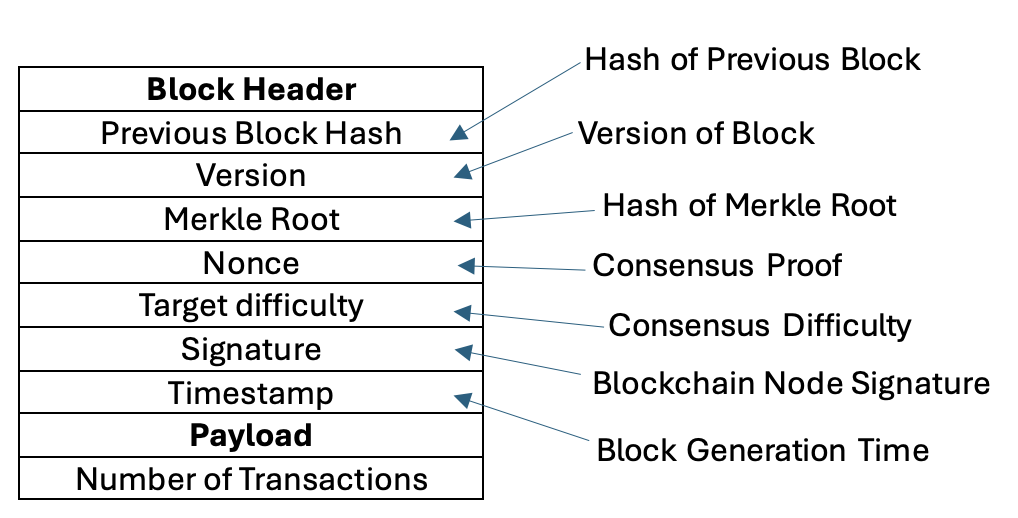}
    \caption{Block format.}
    \label{Fig2}
\end{figure}

In an IoT environment, all the devices are connected to the internet to share their data. Due to the sharing of large amounts of data over the internet, the IoT environment faces security challenges. To ensure the legitimacy of the IoT nodes, we perform authentication. Initially, all the IoT devices and users register their information, such as PUF, Device ID $(Dv-ID)$, User $ID(Ur-ID)$, and MAC address to the key generator (KG) integrated with blockchain. The user provides all four parameters, whereas the IoT devices provide three parameters except $Ur-ID$. The KG then provides a secret key $(K_{pr})$ to the respective devices and users by utilizing DAA, which is stored in the blockchain. Once the registration is over, the respective IoT devices and users become a part of the network. The purpose of DAA's purpose is to provide authentication with less time consumption. During packet transmission and request generation, the devices and users generate a tag $(k_{tg})$ and encrypt the $k_{tg}$ using the encryption key $EN(k_{tg})$. The generation of tags provides more security during authentication. The transaction consisting of $(K_{pr})$, $EN(k_{tg})$, key lifetime $(k_L)$, and current timestamp along with the message and request are submitted to the blockchain. The blockchain verifies the transaction by using the consensus mechanism. During consensus, the blockchain nodes verify whether the submitted credentials are the same as those of the registered ones. If the consensus is achieved and the transaction is verified, the new block is mined, and the respective transaction is recorded in the blockchain. By doing so, the legitimacy of the IoT devices and users is verified before taking part in transmission. Only authenticated IoT devices can send the data through a 5G gateway. The Pseudo code for the proposed blockchain-based authentication process is provided in Algorithm~\ref{Algorithm 1: Authentication Process}. The structure of the generated block in the blockchain is illustrated in Fig.~\ref{Fig2} The description of each field is provided in the figure.

\begin{algorithm}[h]
\SetAlgoLined
\KwIn{$K_{pr}$, message packet/request, time stamp, $k_L$} 
\KwResult{Authentication} 
\textbf{Begin} \\
Generate a tag $k_{tg} \in R_N$  \\
Encrypt $k_{tg}$  $\rightarrow$ EN$(k_{tg})$  \\
Generate transaction T.Authenticate to blockchain  \\ 

\ForEach{T.Authenticate}{
    \If{$k_L < $ time stamp}{
        Verify T.Authenticate through consensus \\
        \eIf{res.consensus = 0}{
            Drop message/request // Illegitimate
        }{
            Accept message/request
        }
    }
    \Else{
        Drop transaction
    }
}
\textbf{End}
\caption{Authentication Process}
\label{Algorithm 1: Authentication Process}
\end{algorithm}

\subsection{Intrusion detection System}

The intrusion detection system is used to detect attacks which are performed through compromised devices in the environment. This improves the security of the IoT environment. In our work, we perform two types of intrusion detection, such as 

\subsubsection{Signature-based intrusion detection}

\begin{figure*}
    \centering
    \includegraphics[height=.57\textwidth]{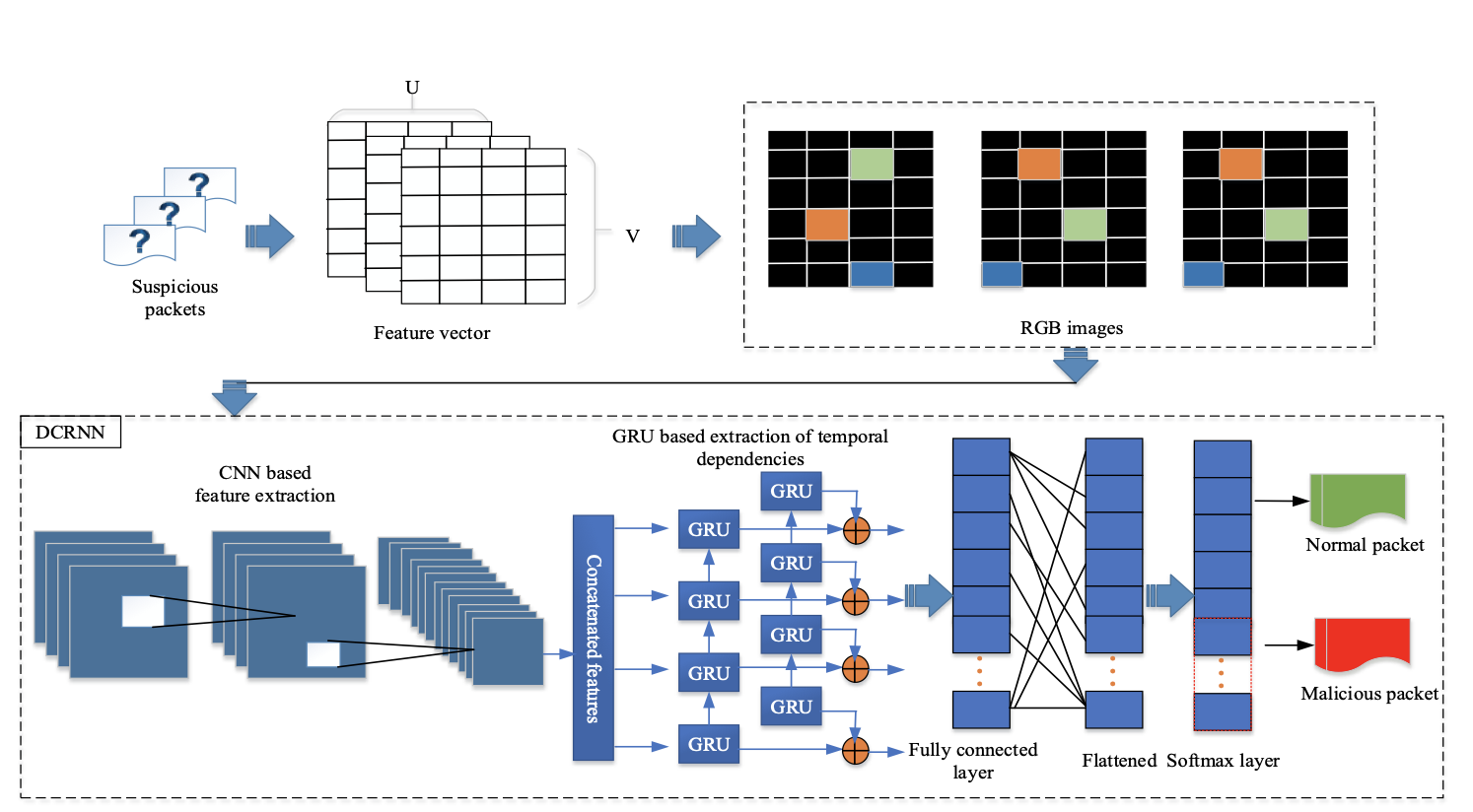}
    \caption{DCRNN-based detection of suspicious packets.}
    \label{Fig3}
\end{figure*}

It is used to detect known attacks that have already been trained and stored in the database. It sends the data to perform intrusion detection. First, we detect signature-based intrusion detection using the IRF. The IRF possesses increased classification accuracy by selecting the important features and optimally constructing the number of trees. The forest $\Theta_0$ is initially grown with $Z_0$ number of trees and feature vector $\varsigma_0$(.). The ranking of features is executed based on the respective weight of the features, which can be computed as,


\begin{equation}
w(k) = \frac{\sum_{\forall T} W^T (k) \cdot \partial^T}{\max\limits_{k} \sum_{\forall T} W^T (k) \cdot \partial^T}
\end{equation}

Where \( W^T(k) \) represents the weight of feature \( k \) with respect to tree \( T \), and \( \partial^T \) denotes the overall weight of tree \( T \). From the ranked list of features, the important features \( h_0 \) are selected and grouped into a pool of important features denoted as \( \Delta \). Similarly, the pool of unimportant features is denoted by \( \Delta' \). The mean and standard deviation of the weights of features in \( \Delta' \) are represented as \( \alpha_0 \) and \( \beta_0 \), respectively. Let \( Rf_0 \) be the number of features in \( \Delta' \) whose weights are less than \( (\alpha_0 - 2\beta_0) \). These features \( Rf_0 \) are further removed from \( \Delta' \) to form a refined feature set \( \varsigma_1(\cdot) = \varsigma_0(\cdot) - Rf_0 \). The condition for including a feature \( k \) from \( \Delta' \) into \( \Delta \) can then be formulated as:

\begin{equation}
w(k) \geq \min_{j \in \Delta} w(j), \quad k \in \Delta'
\end{equation}

If a feature is termed an important feature, it cannot be removed. The number of important and unimportant features can be expressed as $h$ and $g$ from which the variation of important and unimportant features can be computed as:

\begin{equation}
\delta h = \Delta_{n+1} - \Delta_n
\end{equation}
and
\begin{equation}
\delta g = \Delta'_{n+1} - \Delta'_n
\end{equation}

The number of trees to be added to the existing number of trees in the forest in order to improve the accuracy can be computed as:

\begin{equation}
|\delta Z| \leq \left| \frac{l (p_u \delta h + p_g \delta g)}{g} \right|
\end{equation}
where \( l \) is expressed as:

\begin{equation}
l = Z M_{av} P^{M_{av}-1} (1 - P^{M_{av}})^{Z-1}
\end{equation}
where $P$ < 1, $M_{av}$ is the average number of nodes in a tree, and  $p_u$, $p_g$ represent the probability of good split concerning $u$ and $g$. The IRF classifies the data into three classes: normal, malicious, and suspicious. This is done by computing the similarity between the features of the incoming packet and the feature set of known attacks. From the classified packets, the malicious packets are denied, and suspicious packets are sent to the next phase of intrusion detection. The Pseudocode for the proposed IRF-based SIDS is provided in Algorithm~\ref{Algorithm 2: Improved Random Forest Algorithm}.

\begin{algorithm}[h]
\SetAlgoLined
\textbf{Begin} \\ 
\textbf{\textit{Initialization:}} $\Phi_0$, $Z_0$, $\varsigma_0(\cdot)$, and number of passes $n$\\
Compute feature weight using Eq.~(1) and perform feature ranking \\
Select $h_0$ features and group into $\Delta$ \\
Group the rest of the features $g_0$ into $\Delta'$ \\
\While{$g_n \geq f$}{
  Compute $\alpha_n$, $\beta_n$ of features in $\Delta_n'$ \\
  Compute $Rf_n$ and remove it from $\Delta_n'$ \\
  Perform Eq.~(2) and select a set of features $S_n$ in $\Delta_n'$ \\
  Move $S_n$ from $\Delta_n'$ to $\Delta_n$ \\
  Compute $\varsigma_{n+1}(\cdot) = \varsigma_n(\cdot) - Rf_n$ \\
  Compute $\Delta_{n+1} = \Delta_n + S_n$ and $\Delta_{n+1}' = \Delta_n' + S_n$ \\
  Compute $\delta h$, $\delta g$ using Eq.~(3) \\
  Perform Eq.~(4) to determine the value of $\delta Z$ \\
  Compute $Z_{n+1} = Z_n + \delta Z$ \\
  Grow forest $\Phi_{n+1}$, $Z_{n+1}$, and $\varsigma_{n+1}(\cdot)$ \\
  $n = n + 1$
}
\textbf{End while} \\
\textbf{End}
\caption{Improved Random Forest Algorithm}
\label{alg:improved-rf}
\end{algorithm}

\subsubsection{Anomaly-based intrusion detection}

It is used to identify unknown attacks in the IoT environment. The suspicious packets are classified as normal or malicious using DCRNN by considering spatial and temporal features. The proposed neural network possesses a series connection of a CNN and a Gated Recurrent Unit (GRU). The CNN acts as a feature extractor that extracts all the packets' features through the convolution and pooling layers. The extracted features are then provided to the GRU to analyze the temporal dependency between the features to achieve improved classification accuracy. Initially, the packets' local features can be extracted and converted into feature vectors. These vectors are then rescaled from 0 to 255 to get converted into images of $U$ × $V$ pixels where $U$ represent the number of columns, and $V$ represents the number of rows, respectively. The CNN model utilized in our approach extracts the features from the images as formulated below: 

\begin{equation}
I_l' = \frac{I_l - k r + 2Q}{S K} + 1
\end{equation}
where $I_l$ represents the input image length, $Q$ refers to the number of zeros, and $Kr$ refers to the size of the kernel. The stride of the kernel is represented as $SK$. The output of the CNN is led to the GRU, which possesses two gates, namely, the update gate and the reset gate, that work as follows:

\begin{equation}
\begin{aligned}
    c_t &= \lambda \left( w_c \cdot [I_{t-1}, y_t] + y_t \right),   \\
    b_t &= \lambda \left( w_b \cdot [I_{t-1}, y_t] \right),         \\  
    I_t' &= \tanh \left( w \cdot [b_t \cdot I_{t-1}, y_t] \right),  \\
    I_t &= (1 - c_t) \cdot I_{t-1} + c_t \cdot I_t'.   
\end{aligned}
\end{equation}

where 

\begin{equation}
\lambda = \frac{1}{1 + e^{-t}}
\end{equation}

\begin{equation}
\tanh(t) = \frac{1 - e^{-2t}}{1 + e^{-2t}}    
\end{equation}
where the input features are represented by $y$. The $I_t$, $c_t$ represent the update parameters respectively. The weight function is represented as $w$, and the activation function utilized for the flow of sequences is represented as tanh and $\lambda$, respectively. The relationship between both spatial and temporal features is determined with the help of sequence vectors, which are fed into the fully connected and softmax layers for the classification of the packet. If the packet is classified as malicious, then it will be dropped, and the normal packets will be sent to the cloud. Fig.~\ref{Fig3} shows the DCRNN-based detection of suspicious packets.

\subsection{Trust Value Evaluation and Virtual Honeypot Deployment}

\begin{figure*}
    \centering
    \includegraphics[width=.75\textwidth]{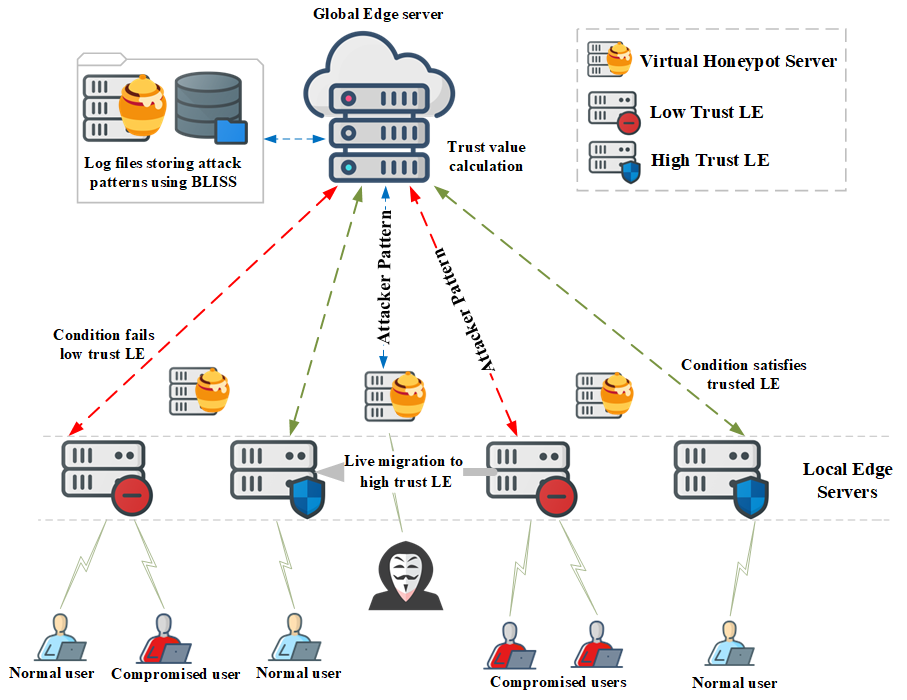} 
    \caption{Virtual Honeypot Deployment and Live Migration}
    \label{Fig4}
\end{figure*}

In this section, we perform a honeypot-based attack prevention mechanism based on trust computation for efficient service provisioning in the network. Fig.~\ref{Fig4} illustrates the trust-based migration of services and virtual honeypot deployment for intrusion prevention. 


\subsubsection{Trust value calculation}

In this model, we proposed two types of edge servers: a global edge server and a local edge server. The global edge server has the responsibility to manage the local edge servers. The user service requests are collected by the local edge server. Then, we calculate the trust value for the edge server as follows:

\begin{equation}
     \tau (LE) = In_\tau \pm (1- In_\tau) X \vartheta
\end{equation}

where, $In_\tau$ and $\vartheta$ denote the initial trust value and degree of change of trust value of the edge servers, respectively. Based on the change in the degree of trust value, the trust value will be dynamically increased or decreased. 

\subsubsection{Live migration}

After calculating the trust value $\tau(LE)$ of the local edge servers, the global edge server computes the threshold value $\tau_{TH}(LE)$ based on the average of all trust values of the local edge servers. The nodes that possess $\tau(LE) \geq \tau_{TH}(LE)$ are classified as high-trusted local servers, denoted as $H(\tau(LE))$, while nodes with $\tau(LE) < \tau_{TH}(LE)$ are classified as low-trusted local edge servers, denoted as $L(\tau(LE))$. 
The $L(\tau(LE))$ nodes are vulnerable to attacks, and to ensure uninterrupted service to users, we implement live migration using a moving target defence strategy. This approach enables the migration of running services from $L(\tau(LE))$ to a suitable $H(\tau(LE))$. The selected $H(\tau(LE))$ must have a system configuration similar to that of $L(\tau(LE))$. 
To determine an appropriate $H(\tau(LE))$ with the same system configuration as $L(\tau(LE))$, we propose a Heap-Based Optimization (HBO) method that considers CPU resource availability, trust value, and system load. The definitions of the input parameters are as follows:

\begin{itemize}
\item CPU Resource availability: CPU resource availability $RA_{CPU}$  is defined as the summation of the number of currently running tasks running on the edge server, and the tasks are provided to the available server based on the user request. It is denoted as:

\begin{equation}
     RA_{CPU} = \sum_{i=1}^{n} E_i T_i
\end{equation}


\item Load: The load $(L)$ represents the number of user tasks waiting $(T_w)$ to be processed by the edge server, relative to the available CPU resources $(RA_{CPU})$.

\begin{equation}
     L = \frac{RA_{CPU}}{T_w}
\end{equation}

\end{itemize}

Based on the above input descriptions, Heap-Based Optimization (HBO) is performed. The HBO is an optimization algorithm for finding the optimal path shown in Algorithm~\ref{Algorithm 3: Heap-based Optimization}. Here, the objective of HBO is to find the H $(\tau(LE))$ based on maximum CPU resource, maximum trust value and minimum load. Construction of an optimal heap needs a Search Agent (SA), which includes Fitness Value (FV), which is the heap nodes' key, and index (I), which has the heap nodes' value. The SA is calculated based on the no. of LEs which are present in the IoT environment, and it is denoted as:

\begin{equation}
\begin{bmatrix}
SA_1^{DIM} & SA_2^{DIM} & \cdots & SA_M^{DIM} \\
\vdots & \vdots & \ddots & \vdots \\
SA_N^{DIM} & SA_{N-1}^{DIM} & \cdots & SA_M^{DIM}
\end{bmatrix}
\end{equation}
where $N$ denotes the number of $SA$, and DIM denotes the corresponding dimensions. A $FV$ is calculated based on $RA_{CPU}$, $\tau(LE)$, $L$ for each $LE$ servers which is denoted as:

\begin{equation}
FV=\{FV(LE_1),FV(LE_2),\cdots, FV(LE_n)\}
\end{equation}

Based on the number of search agents ($SA$), the fitness values ($FV$), and the corresponding index values ($I$) are calculated. The index ($I$) represents the value of the heap node based on CPU resource availability ($RA_{CPU}$), trust level ($\tau(LE)$), and system load ($L$). The average fitness value ($AVG_{FV}$) is computed based on the objective conditions. If any node satisfies all the objective conditions during the computation of $AVG_{FV}$, it is assigned as the root node ($R_n$) at depth level $d=0$. Subsequently, for corresponding depth levels $d = 1, 2, 3, \dots, n$, branch nodes ($B_n$) are assigned dynamically based on system conditions (i.e., trust level, CPU availability, and load). For root node calculation, we have:

\begin{equation}
R_n = 
\begin{cases}
\text{high } RA_{\text{CPU}}, & \text{if } d = 0
\end{cases}
\end{equation}

For branch node allocation at increasing depths:
\begin{equation}
B_n(d) =
\begin{cases}
H(\tau(LE)) \geq \tau_{TH}(LE), & d = 1 \\
L_{\text{low}}, & d = 2 \\
B_n(d-1) > B_n(d), & d = 3, 4, \dots, n
\end{cases}
\end{equation}

For corresponding depth levels $d = 1, 2, 3, \dots, n$, branch nodes ($B_n$) are assigned progressively in a hierarchical manner according to trust level, load, and CPU availability. The root node ($R_n$) is initialized at $d = 0$, and as depth increases, branch nodes are dynamically selected based on system conditions. To maintain hierarchical ordering:

\begin{equation}
B_n(d) < R_n, \quad \text{for all} \quad d \geq 1
\end{equation}

For building a heap $R_n$ branch nodes based on d levels are denoted as:

\begin{equation}
H(LE) =  \begin{cases}
R_n \geq highRA_{CPU}, &  \tau(LE) , L,d= 0 \\
B_n(1)<R_n, &  d= 1 \\
B_n(n)<B_n(n-1), &  d= n
\end{cases}
\end{equation}

\begin{algorithm}[h]
\SetAlgoLined
\KwIn{$RA_{CPU}$, $\tau(LE)$, $L$} 
\KwResult{$R_n$, $high\ RA_{CPU}$, $H(\tau(LE)) \geq \tau_{TH}(LE)$, low $L$} 
\textbf{Begin}

Calculate heap nodes based on $RA_{CPU}$, $\tau(LE)$, and $L$;\
Compute average fitness value $AVG_{FV}$ based on objectives;\

\ForEach{node in nodes}{
\If{node satisfies all objective conditions for $AVG_{FV}$}{
Assign node as Root node ($R_n$) at depth level $d = 0$;\
}

Assign branch nodes $B_n$ dynamically for depth levels $d=1,2,3,...,n$ based on trust, load, CPU;\

\ForEach{transaction request}{
\If{timestamp is valid}{
Validate request via consensus;\
\eIf{consensus result = valid}{
Accept transaction;
}{
Drop transaction (illegitimate request);\
}
}
\Else{
Drop transaction (invalid timestamp);
}
}
}
\textbf{End}
\caption{Heap-based Optimization}
\label{algo:heap_optimization}
\end{algorithm}

After heap construction based on $AVG_{FV}$, the services from $L (\tau(LE))$ are migrated to heaps of $H (LE)$ based on heaps arranged hierarchically.

\subsubsection{Virtual Honeypot Deployment}

\begin{table}[H]
    \centering
    \caption{SYSTEM CONFIGURATION}
    \begin{tabular}{|c|p{13mm}|p{30mm}|}
        \hline
        \multirow{4}{*}{Hardware Specifications} & \textbf{Hard Disk} & 60 GB \\ \cline{2-3}
                                           & \textbf{RAM}       & 8 GB  \\ 
                                           \cline{2-3}
         & \textbf{Processor} & Intel (R) Core (TM) i5-4590S CPU @ 3.00 GHz \\ \hline
        \multirow{2}{*}{Software Specifications}                                & \textbf{Operating System} & Ubuntu 14.04 LTS \\ \cline{2-3}
                & \textbf{Network Simulator} & NS3.26 \\ \hline
    \end{tabular}
    \label{SYSTEM CONFIGURATION}
\end{table}

During live migration, virtual honeypots with the same system configuration as the Local Edge Servers (LEs) are deployed to attract attackers and prevent the compromise of LEs. These honeypots detect attack signatures and store them in log files, which are then securely stored in the database. To enhance security, the Bimodal Lattice Signature Scheme (BLISS) algorithm is employed for encryption, as shown in Algorithm~\ref{Algorithm 4: BLISS Signature Generation}. BLISS is a digital signature scheme that uses a public key for verification and a private key for signature generation. The primary objective is to encrypt log files containing attack signatures before storing them in the database. During migration, attackers might attempt to compromise \( H(\tau(LE)) \). To counter this, virtual honeypots \( (HP_1, HP_2, HP_3, \dots, HP_n) \) — configured identically to the LEs — are deployed, misleading attackers into targeting the honeypots instead of the legitimate LEs. Once an attack is detected on a honeypot, its signature is recorded in a log file \( Lf \) and encrypted using the BLISS algorithm. The attack pattern, denoted as \( \omega \), captures various attacker behaviours targeting the honeypot. The verification function \( G \) and signature generation function \( H \) process the attack signature, where \( (S, F) \) represents the generated signature and the challenge vector for computing the hash function. The sample vector \( d \) is derived from an \( n \)-dimensional Gaussian distribution with a standard deviation \( \sigma \), ensuring randomness in signature generation. A random bit \( r_e \in \{0,1\} \) is selected, and a shift vector \( K_c \) is applied to finalize the signature. The resulting \( S \) is encrypted using BLISS and securely stored in the log file.

\begin{equation}
\begin{aligned}
    F &= H(\omega) \quad \text{ // Compute hash of attack pattern} \\
    d &\sim \mathcal{N}_{\sigma}^{n} \quad \text{ // Sample from Gaussian distribution} \\
    r_e &\in \{0,1\} \quad \text{ // Generate random bit} \\
    S &= d + r_e K_c \quad \text{ // Generate attack signature} \\
    Lf &= \text{Encrypt}(S, F) \quad \text{ // Encrypt and store in log file}
\end{aligned}
\end{equation}

\section{RESULTS AND ANALYSIS}

\begin{algorithm}[h]
\SetAlgoLined
\KwIn{Attack pattern $w$, Public key $G \in \mathfrak{O}_{2p}^{n \times m}$, Private key $H \in \mathfrak{O}_{2p}^{n \times m}$, Security parameter $\sigma \in \mathbb{R}$} 
\KwResult{Signature $(S, F)$ of the packet pattern $w$} 
\textbf{Begin} \\
$d \gets$ Sample vector from Gaussian distribution $\mathcal{N}(0,\sigma^2I_n)$\\[4pt]
$F \gets H\left((G \cdot d) \mod 2p,\; w\right)$ \quad // Compute hash-based challenge\\[4pt]
Randomly select bit $re \in \{0,1\}$\\[4pt]
$S \gets d + (-1)^{re} K_c$ \quad // Compute signature response\\[4pt]
Output $(S, F)$ if acceptance criteria are satisfied\\[4pt]
Store $(S, F)$ securely encrypted in log file $Lf$ \\[4pt]

\If{acceptance criteria are not satisfied}{
    Restart signature generation
}
\textbf{End}
\caption{BLISS Signature Generation}
\label{Algorithm 4: BLISS Signature Generation}
\end{algorithm}

This section deals with the simulation of the proposed approach. The performance of this approach is validated by comparing it with existing approaches. This section is divided into four subsections, namely simulation setup, dataset description, comparative analysis, and research summary, which are presented below.

\begin{table}[H]
\caption{SIMULATION CONFIGURATION}
\begin{tabular}{|p{4cm}|p{3cm}|}
\hline
\multicolumn{2}{|c|}{\textbf{Network Parameters}} \\
\hline
Area of simulation& $900 \times 1200$  \\
\hline
Number of IoT user nodes & 50  \\ \hline
Number of IoT device nodes & 50   \\\hline
Number of  edge gateways &6  \\ \hline
Number of  cloud server &1  \\ \hline
Time for simulation & 200s  \\ \hline
Energy at initial stage & 40J \\ \hline
Modules & IPV4, IPV6, MAC  \\ \hline
Number of malicious nodes & 10-30 nodes \\ \hline
Number of 5G base station & 1  \\ \hline
\multicolumn{2}{|c|}{\textbf{Packet Transmission parameters}} \\
\hline
Packets data rate & 500Mbps  \\
\hline
Packets interval & $2^6$, $2^7$, $2^8$, $2^9$, $2^{10}$ bytes  \\ \hline
Number of packets & 1250    \\\hline
Retransmission rate & 10 \\ \hline
Protocol used  & UDP  \\ \hline
\multicolumn{2}{|c|}{\textbf{Network Traffic Parameters}} \\
\hline
BW of the channel& 100KHz  \\
\hline
Queue used & First In, First Out \\ \hline
Traffic type & UDP, TCP, CBR    \\\hline
\multicolumn{2}{|c|}{\textbf{Security Parameters}} \\
\hline
Attack probability ratio& 1:5  \\
\hline
Attacks interval & 2-6 p/sec  \\ \hline
Attacks detected &  93\%  \\\hline
Number of attacks & $\sim8$  \\ \hline
Frequency of attacks   & 15-30p/sec \\ \hline
\multicolumn{2}{|c|}{\textbf{Mobility parameters}} \\
\hline
Country Name or Area Name& ISO ALPHA 2 Code  \\
\hline
Transmission range & 250m  \\ \hline
Model of mobility & Random waypoint model   \\\hline
Nodes mobility & 15.3m/s \\ \hline
\end{tabular}
\label{SIMULATION CONFIGURATION}
\end{table}

\subsection{Simulation Setup}

The simulation tool used for the implementation of the proposed approach is NS3.26, in which IDS detection, AIDS detection, virtual honeypot deployment and attack signature pattern are successfully verified. Table~\ref{SYSTEM CONFIGURATION} shows the system configuration for attaining the simulation. Table~\ref{SIMULATION CONFIGURATION} shows the simulation configuration for the proposed approach. In our proposed approach, the dataset used is CICIDS-2017 for designing an effective attack detection system, which is specially designed for IDS and AIDS. This dataset contains current and old benign (normal traffic) and attacks (anomaly traffic) such as DoS, DDoS, Web-based, Heart bleed, Infiltration, Scan and Bot. Based on parameters such as source and destination ports, source and destination IPs and time stamps, network traffic was analyzed using a CIC flow meter and labelled in a CSV file. At that time of implementation, data can be separated into training data and test data. The training data contains 157,722 packets, among which 137,626 are labelled as normal (benign) and 20,076 are labelled as attacks. The test data contains 66,834 packets. Among that, 54,284 are labelled as normal, and 12,540 are labelled as attacks. The records of the attacks are stored in a CSV file. The features of the dataset are shown in Table~\ref{SIMULATION CONFIGURATION}.

\subsection{Comparative analysis}

The validation of the proposed approach is presented in this sub-section, in which our approach is compared with the existing approaches such as RADDS \cite{wahab2019resource} and GTM-Csec \cite{gill2020gtm} by means of several performance metrics such as accuracy, attack detection rate, false negative rate, precision, recall, ROC curve, memory usage, CPU usage, and execution time.

\subsubsection{Impact of attack detection rate}

The attack detection rate of any system is defined as the rate of detection of attack packets from the overall packets. It must be high so that the system is secure in any environment; if any system has a low attack detection rate, it will lead to severe security threats. 

\begin{figure}
    \centering
    \includegraphics[height=.4\textwidth]{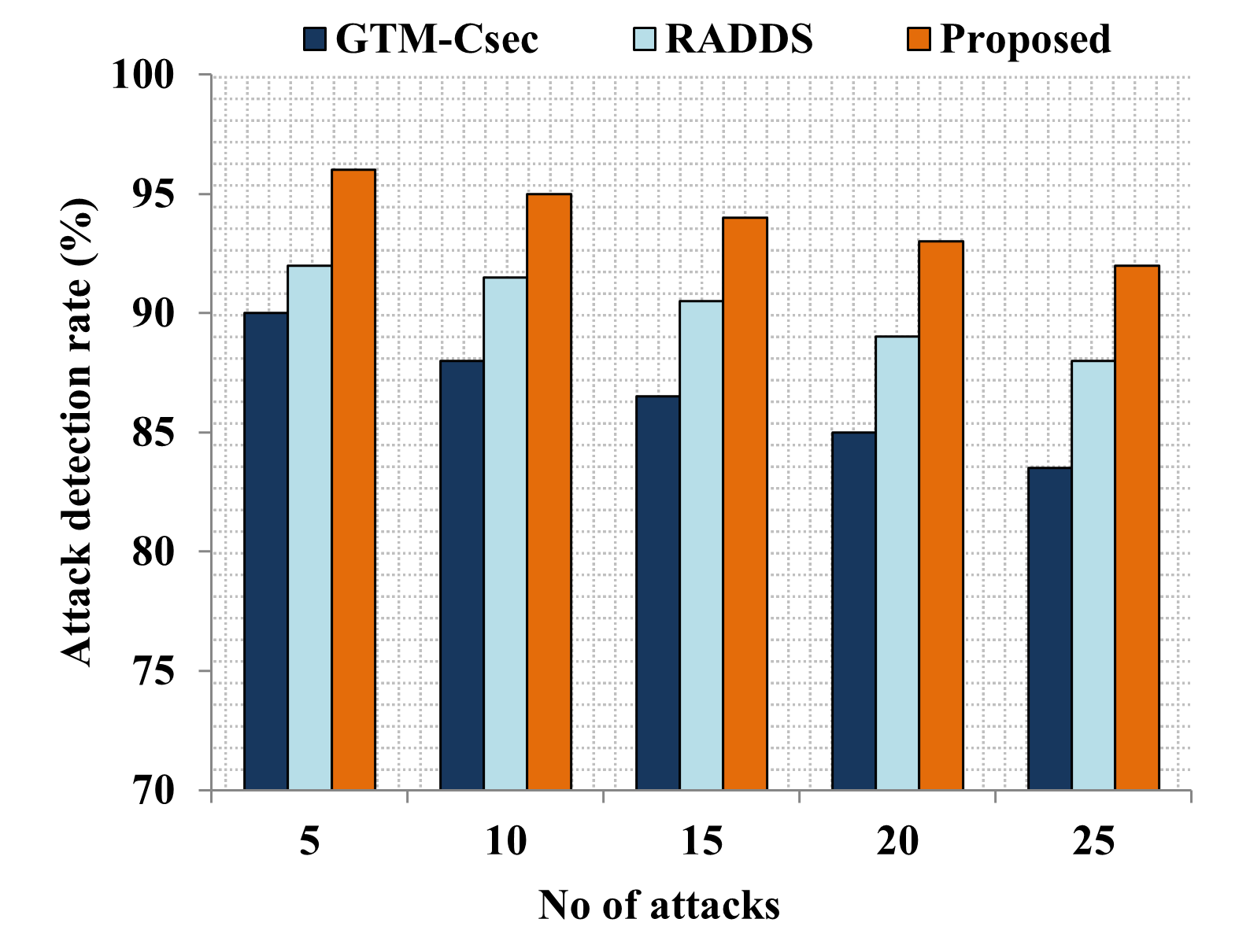}
    \caption{Number of attacks vs. attack detection rate}
    \label{Fig5}
\end{figure}
 
Compared to the existing works in Fig.~\ref{Fig5}, the attack detection rate of the proposed approach is 96.5\%, even though the attacks increased. The deployment of a virtual honeypot collects the pattern of attacks and stores it in the log in the database in an encrypted manner. These attack patterns are further used to train the SIDS model, which helps to detect the attacks earlier, while the existing works use VM techniques and physical honey pot deployment, which have an attack detection rate of 92\% and 89.7\%, and do not focus on the collection of attack patterns. The existing approaches utilized game-theoretic approaches in which they assume that the attack strategies of the attacker are known to the IDS, but in reality, this is not so; this leads to a low attack detection rate. The proposed approach outperforms existing approaches with an increased attack detection rate of 4.5\% to 7.2\%.

 \subsubsection{Impact of accuracy}

\begin{figure}
    \centering
    \includegraphics[height=.36\textwidth]{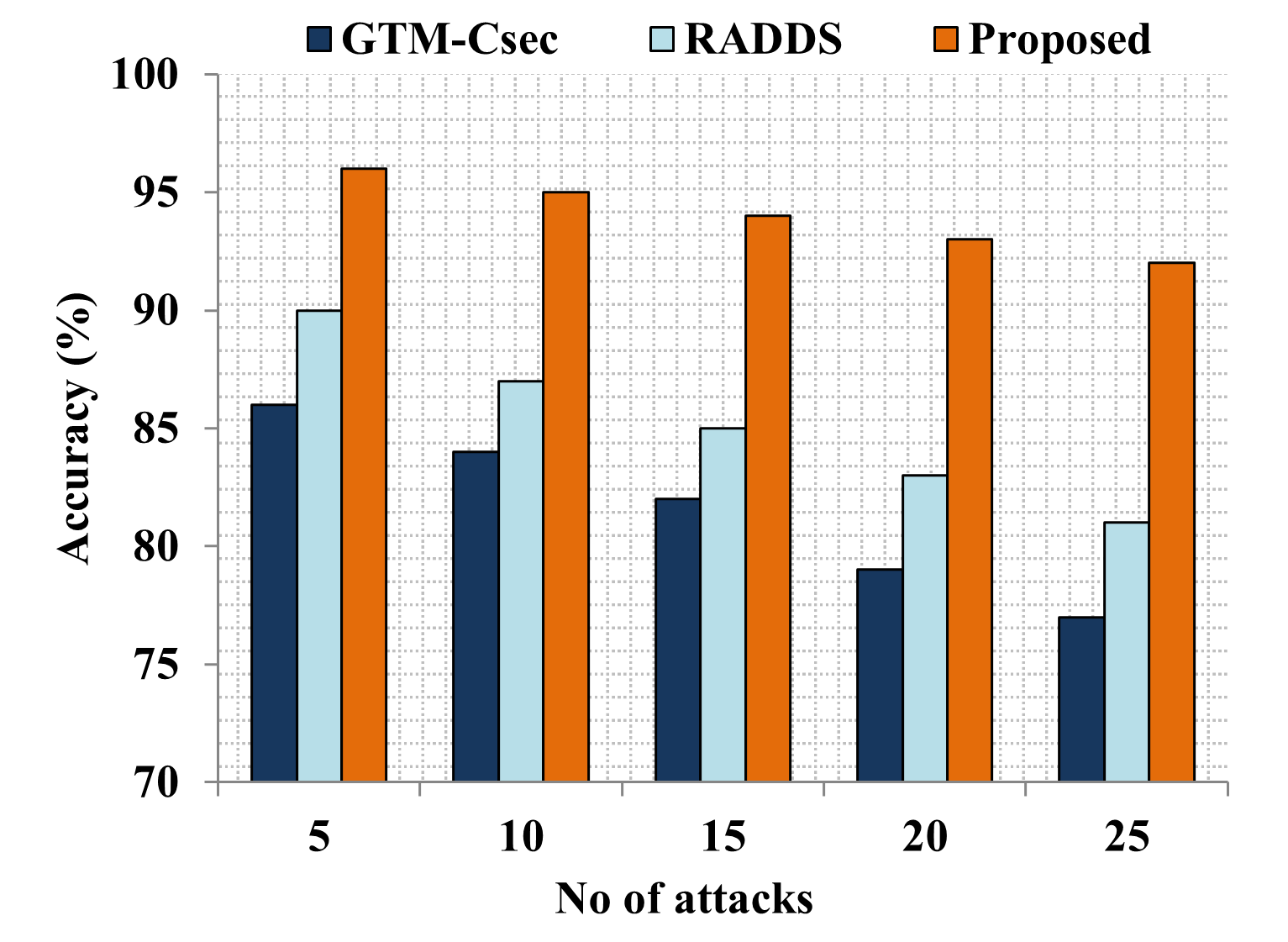}
    \caption{Number of Attacks vs. Accuracy}
    \label{Fig6}
\end{figure}

\begin{figure}
    \centering
    \includegraphics[height=.34\textwidth]{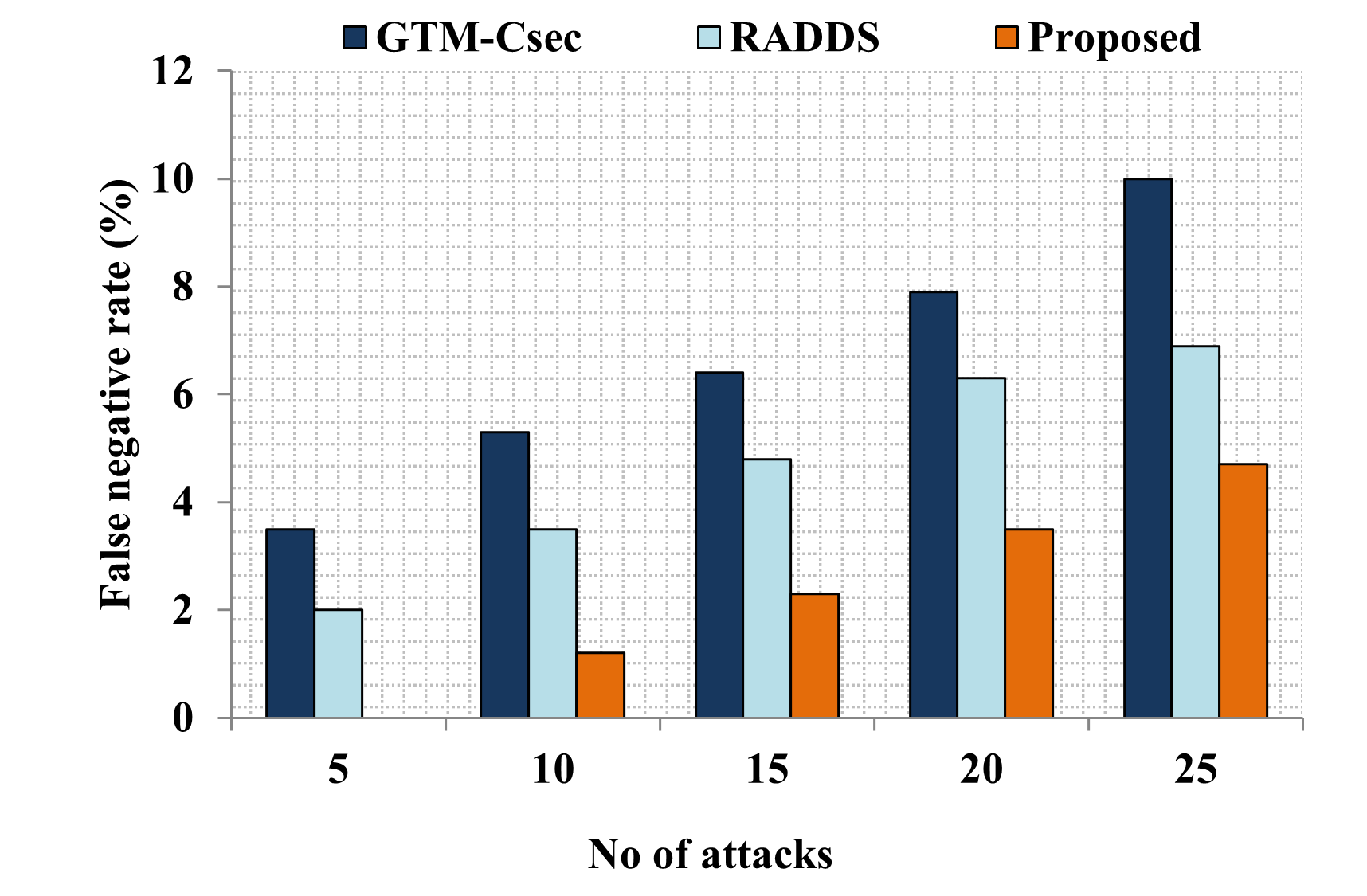}
    \caption{Number of Attacks vs. False Negative Rate}
    \label{Fig7}
\end{figure}

Accuracy is termed the capability of a system to determine the type of attack. The accuracy should be high for any system in terms of attack detection; low accuracy leads to increased risk in the system's environment and causes security threats. The detection accuracy of our proposed model, shown in Fig.~\ref{Fig6}, is 97\%, which is more efficient than the existing works; however, the no. of attacks increases. The DCRNN-based detection of anomalies based on both spatial and temporal features increases the accuracy, while the existing works use a Game-theoretic model for security, which has an accuracy of 89\% and 84\%, and does not focus on the attack pattern collection, which reduces the overall accuracy of these approaches. The proposed model outperforms the existing approaches with an increased accuracy rate of 8\% to 13\%. The False negative rate comparison of the existing and proposed model is shown in Fig.~\ref{Fig7}. Whenever the number of attacks increased, the false negative rate also increased. The false negative rate is just 4.5\% in our proposed method for 25 attacks because of the integrated action of the SIDS and AIDS detection process, which accurately detects both the known and unknown attacks. The existing approaches do not consider the unknown attack strategies, which results in a false negative rate of 7.6\% and 9.7\%, and degrades the performance. The proposed approach outperforms existing approaches with a decreased false negative rate of 3.1\% to 5.2\%.

 \subsubsection{Impact of false negative rate}

The false negative rate is defined as the number of attacks that were falsely determined; a system is a good system when it has a low false negative rate. 

 \subsubsection{Impact of precision}

\begin{figure}
    \centering
    \includegraphics[height=.36\textwidth]{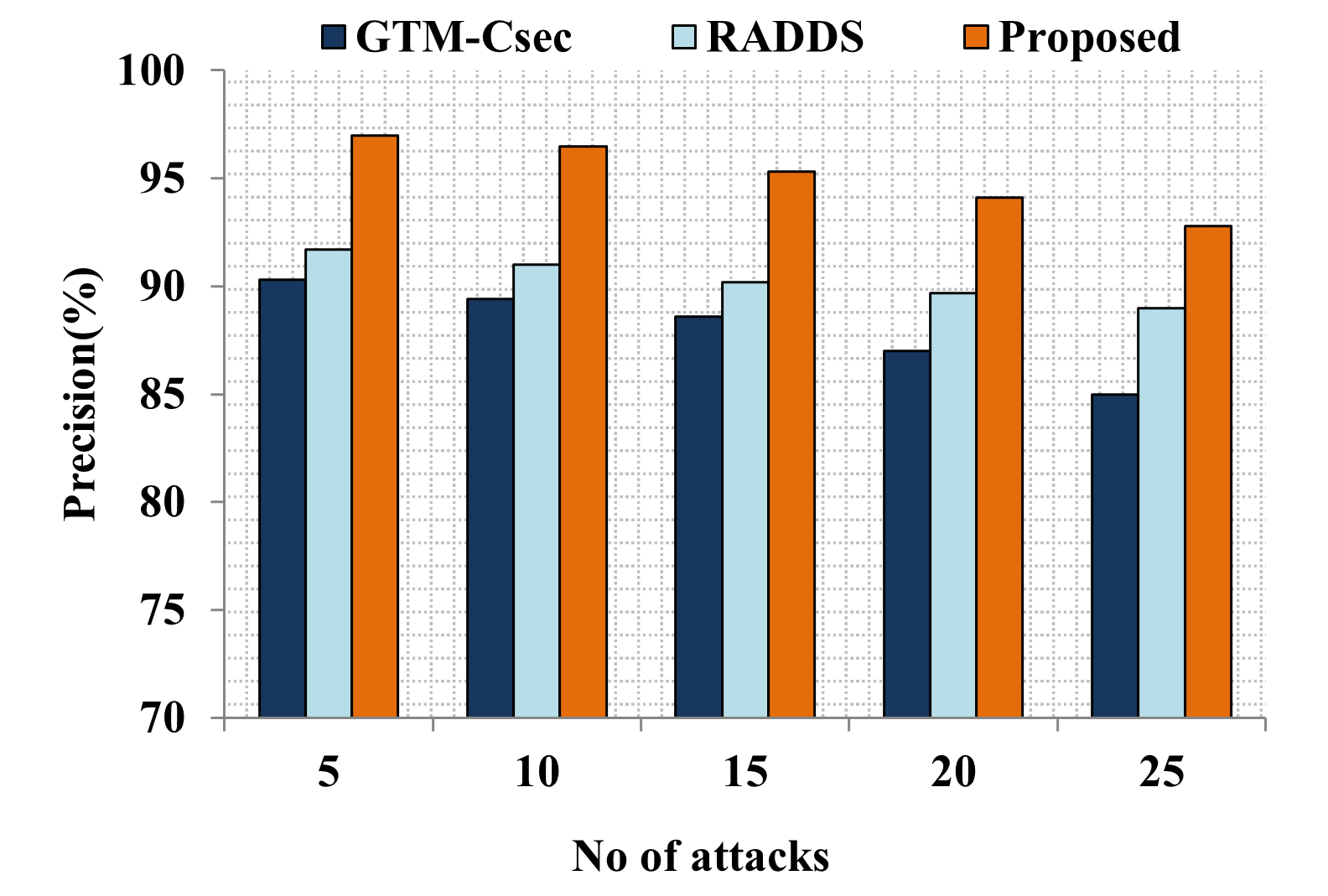}
    \caption{Number of Attacks vs. Precision}
    \label{Fig8}
\end{figure}

Precision is defined as the trueness of a system; if a system has a high trueness value, it has a high precision rate and increased accuracy. Fig.~\ref{Fig8} shows the comparison of the precision of the proposed method and existing approaches with respect to a number of attacks. The precision of the proposed approach is 97\%, whereas the existing works have precision rates of 92.7\% and 90\%, respectively. The trueness of the system in the proposed method is improved by IDS techniques and virtual honeypot deployment for attack pattern collection, while the existing works assumed knowledge of the attack patterns, which decreases the precision rate. The proposed approach outperforms existing approaches with an increased precision rate of 5.2\% to 7\%.

\begin{figure}
    \centering
    \includegraphics[height=.34\textwidth]{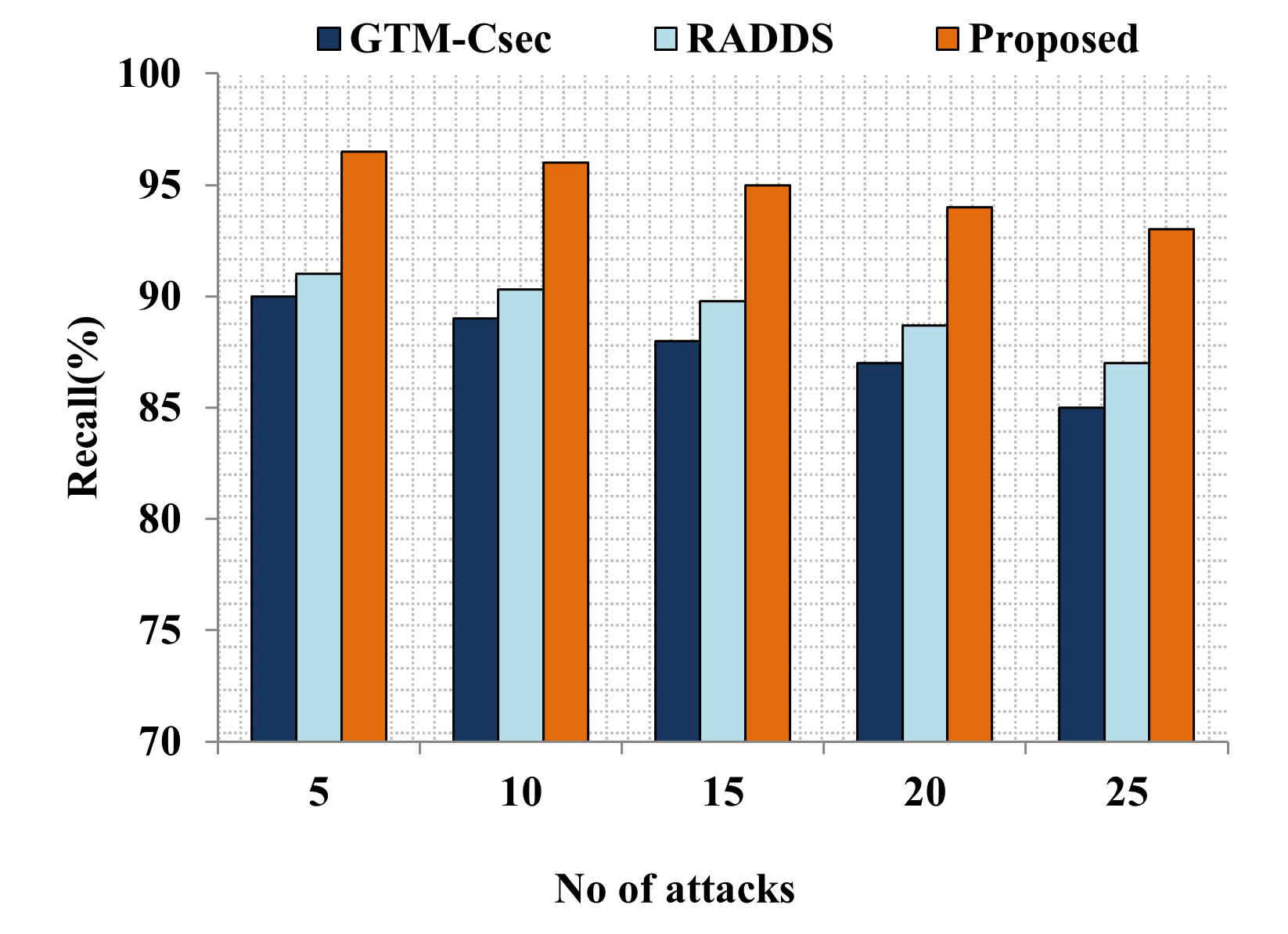}
    \caption{Number of Attacks vs. Recall}
    \label{Fig9}
\end{figure}

 \subsubsection{Impact of recall}

Recall rate is defined as the detection of falseness of the system with high accuracy; if a system has a high recall rate, it would have high overall accuracy. The comparison of the proposed method, which has a recall rate of 96\%, and existing works, which have a recall rate of 92\% and 90.5\%, respectively, with respect to the number of attacks, is shown in Fig.~\ref{Fig9}. The high recall rate is due to the improved detection techniques, which consider both IDS and anomaly IDS in our proposed method, while the existing works did not focus on two types of IDS, leading to a lower recall rate and less overall accuracy. The proposed work improves the recall rate of 4\% to 5.5\% in existing works.
 
\subsubsection{Impact of memory usage}

\begin{figure}
    \centering
    \includegraphics[height=.32\textwidth]{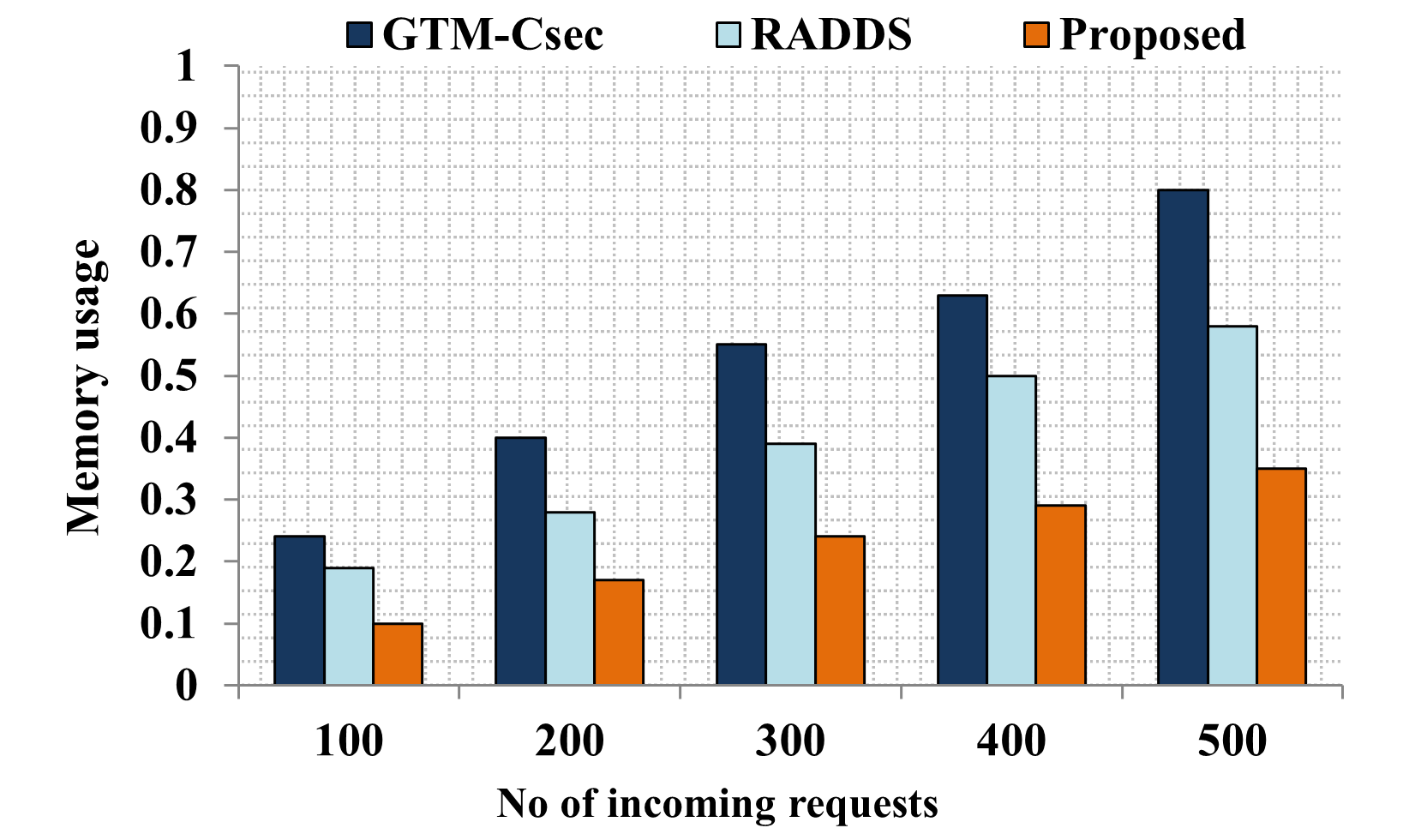}
    \caption{Number of incoming requests vs. memory usage.}
    \label{Fig10}
\end{figure}

Memory usage depends on the amount of memory used for a task to complete. If more memory is utilized, it leads to high energy consumption and overload to the system. The memory usage comparison of the proposed method and existing works is shown in Fig.~\ref{Fig10}. Whenever the number of user requests increases, memory usage increases. The memory usage of the proposed approach is 0.35 due to the migration of services from less trusted to highly trusted by considering the load, which reduces the overhead and energy consumption, while the other existing works have high memory usage of 0.8 and 0.56. This is due to the lack of focus on the load, which increases the memory usage and overhead. The proposed approach outperforms existing approaches with decreased memory usage of 0.45 to 0.21.

 \subsubsection{Impact of CPU usage}

\begin{figure}
    \centering
    \includegraphics[height=.35\textwidth]{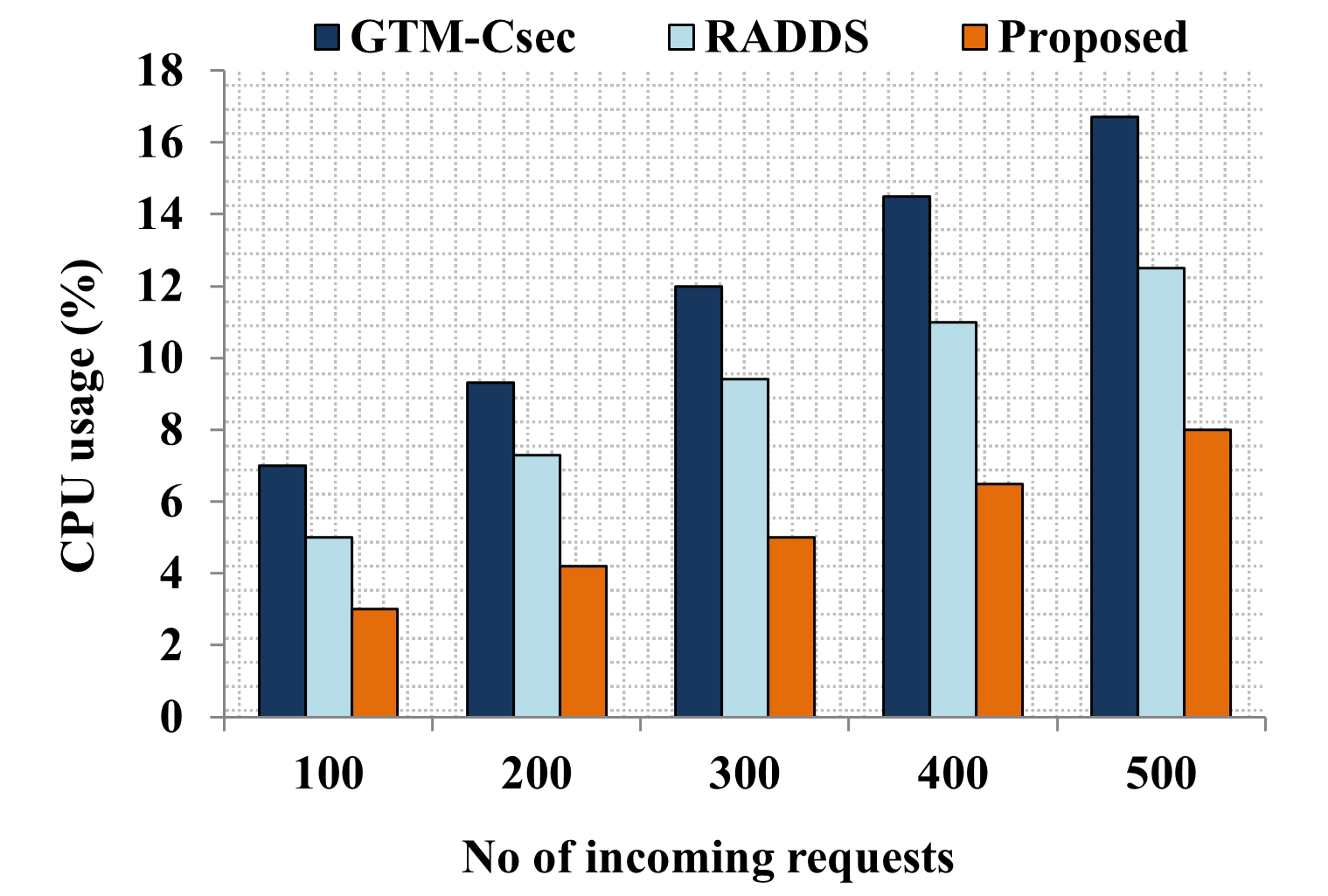}
    \caption{Number of incoming requests vs. CPU usage.}
    \label{Fig11}
\end{figure}

CPU usage is defined as the amount of resources consumed to complete tasks; an efficient system must have low CPU usage and complete more tasks. If a system uses more CPU, it would be termed as an inefficient system and would degrade the overall performance of that system. Fig.~\ref{Fig11} shows a comparison of the CPU usage to 6.7\% with respect to the number of incoming requests with state-of-the-art works. The results show that the CPU usage of the proposed method is less with an increase in requests; however, the existing works have 12\% and 15.6\%, which did not consider the decentralized edge nodes for management, leading to high CPU usage and high resource consumption. The proposed work reduces the CPU usage to 5.5\% and 8.8\% from existing works.

 \subsubsection{Impact of execution time}

\begin{figure}
    \centering
    \includegraphics[height=.31\textwidth]{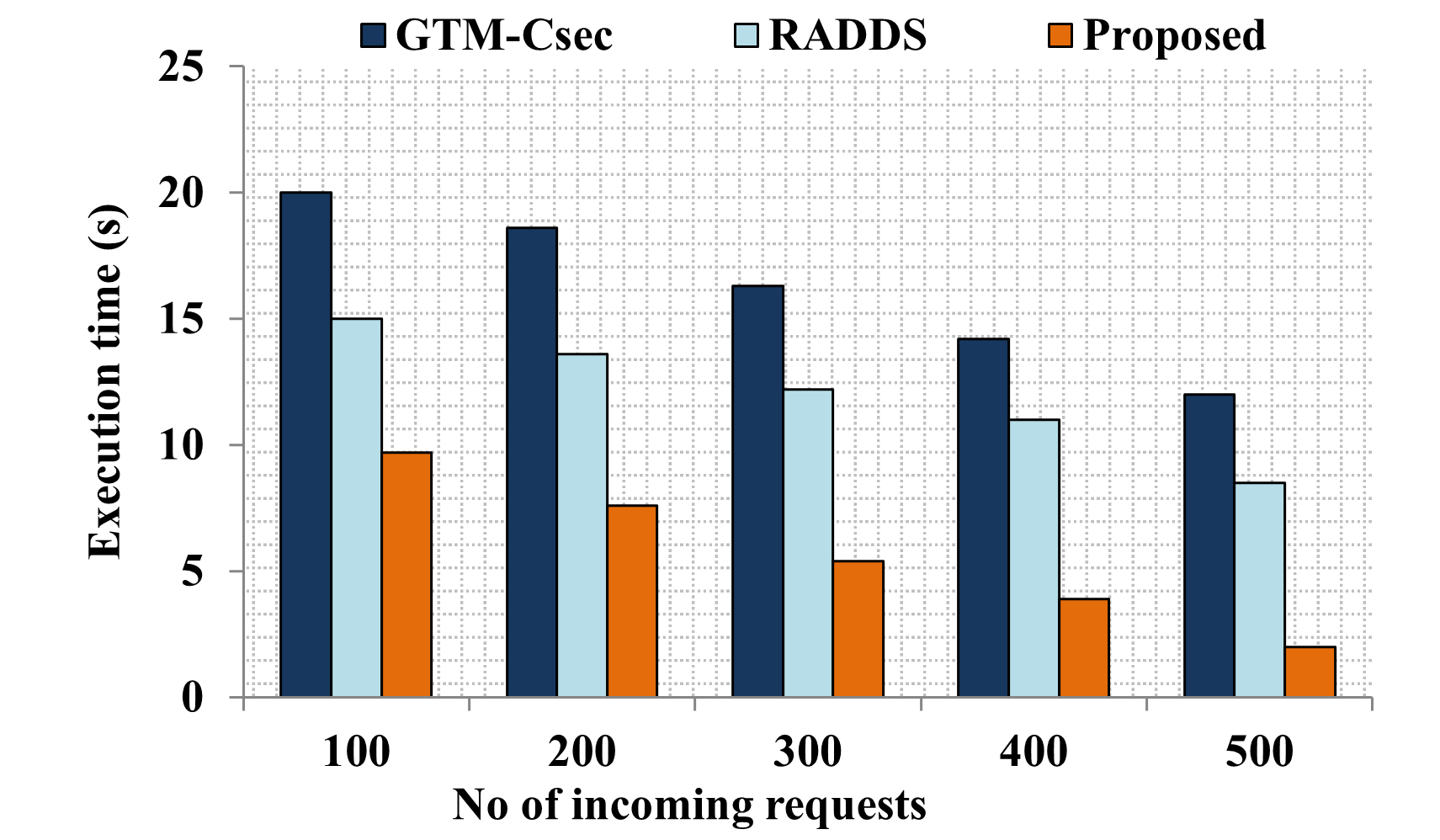}
    \caption{Number of incoming requests vs. execution time.}
    \label{Fig12}
\end{figure}

Execution time must be less for any system to complete a task, which reduces the time complexity and increases the efficiency; if a system has a high execution time, it will lead to high time complexity and performance degradation. A comparison of the proposed method and existing works is shown in Fig.~\ref{Fig12} with respect to the number of incoming requests and execution time. The results show that the execution time of any task is 2.5s even though the incoming requests are increased. This is due to the use of edge computing servers, while the existing works have 9.2s and 14.2s and do not consider the edge servers for network management, which increases the execution time and increases latency. Our approach outperforms existing approaches with a low execution time of 12.3s to 7.3s.

 \subsubsection{Impact of ROC curve}

\begin{figure}
    \centering
    \includegraphics[height=.35\textwidth]{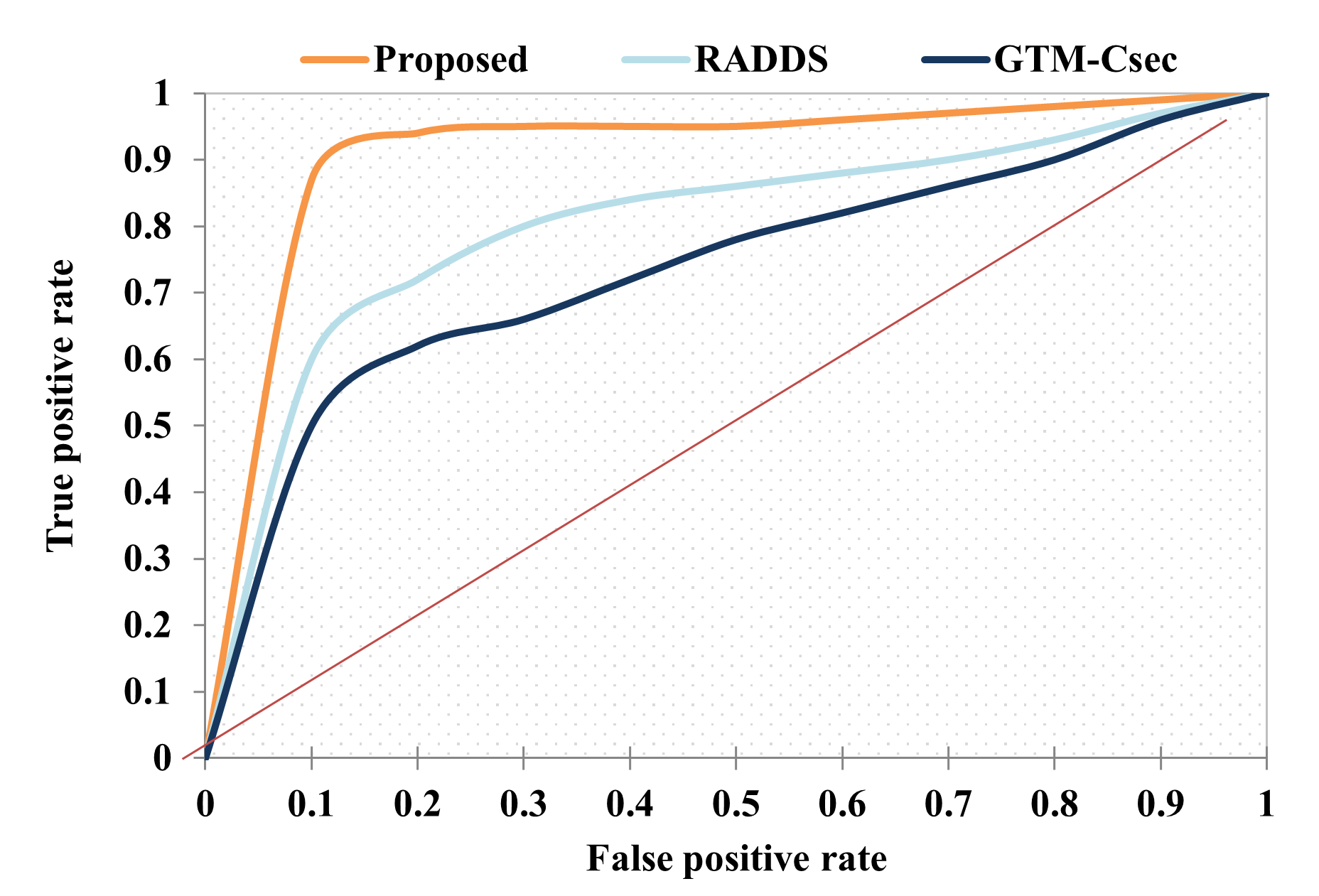}
    \caption{ROC curve.}
    \label{Fig13}
\end{figure}

The Receiver Operating Characteristics (ROC) curve is the classification of the true positive rate and false positive rate of any system at different thresholds. Fig.~\ref{Fig13} shows that the ROC comparison of existing methods and the proposed method has reduced the false positive rate and increased the true positive rate than existing works because of the collection of attack patterns, resulting in an increased detection rate. The existing works assume that the attack strategies of the attacker are known to the IDS, and unknown IDS are not considered during authentication, which increases the false positive rate and decreases the true positive rate.

 \subsubsection{Security Analysis}

 The security of service provisioning in the 5G-IoT environment is hindered by various cyberattacks prevailing in the network. These attacks pose a serious threat to the legitimacy and integrity of the entities in the network. The following are a few of the serious attacks mitigated by the proposed approach.

 \begin{itemize}
     \item Fuzzing attacks – In this type of attack, the attacker attempts to steal the credentials of the devices and users in the network. The proposed approach mitigates these attacks by utilizing blockchain technology in which the credentials are stored in a hashed format. The DAA strengthens the mitigation of fuzzing attacks by employing a tweakable block cipher (Deoxys-TBC) to encrypt and authenticate credentials before they are stored on the blockchain. Unlike conventional hashing techniques, Deoxys-TBC provides enhanced security by introducing tweak-based encryption, ensuring that each authentication request is uniquely processed with a cryptographic nonce, making brute-force credential guessing infeasible. Additionally, mutual authentication mechanisms verify both the requester and the verifier, preventing unauthorized access to stored credentials.

     \item DDoS attacks – The DDoS attackers forward a massive number of packets or user requests with the intention of shutting down edge gateways or the cloud server. The proposed approach performs integrated IDS to determine the nature of the packets, and on the user side, the blockchain verifies the timestamp of the requests. Additionally, the system employs a rate-limiting mechanism and anomaly-based traffic filtering to detect and mitigate high-volume malicious requests. The IDS utilizes machine learning-based pattern recognition to distinguish between legitimate and malicious traffic, reducing false positives while ensuring uninterrupted service availability. Furthermore, blockchain’s decentralized validation prevents single points of failure, as distributed nodes collaboratively verify request authenticity, making it significantly harder for attackers to overwhelm the system. 

     \item Impersonation attack – The attacker attempts to imitate a legitimate user to get into the network to perform various attacks. This attack is mitigated by performing blockchain-based authentication in which the tag generated by the authentication algorithm is verified. To further strengthen security, multi-factor authentication (MFA) can be integrated as an additional layer of defence. By combining blockchain authentication with a secondary verification factor—such as a one-time password (OTP) or biometric authentication—unauthorized access attempts can be significantly reduced. Implementing MFA alongside the existing authentication mechanism would ensure that even if one credential is compromised, attackers cannot easily impersonate legitimate users.

     \item Replay attack – The attacker injects previously captured legitimate packets into the network at regular intervals to exhaust system resources and disrupt normal operations. The proposed approach mitigates these attacks by considering spatio-temporal features for attack detection, which analyze the timing and spatial distribution of incoming packets to distinguish between legitimate transmissions and replayed attacks. By leveraging historical traffic patterns and anomaly detection, the system effectively flags repeated identical requests, preventing attackers from overwhelming network resources. Additionally, blockchain’s timestamping mechanism ensures the uniqueness of transactions, making it difficult for attackers to reuse previously recorded packets without detection.

     
 \end{itemize}

 \subsubsection{Analysis Summary}

In this section, the discussion about the performance of the proposed approach is presented in an elaborative manner. The numerical analysis of the performance of the proposed approach is provided in the table. From the table, it is clear that the proposed approach possesses an increased attack detection rate and accuracy; this is due to the integration of IRF-based SIDS and DCRNN-based AIDS, considering both attack patterns and spatiotemporal features of packets. The updation of attack strategies for the newly caused attacks with the help of the deployment of virtual honeypot nodes also contributes to the increased attack detection rate. The reduced false negative rate and increased precision and recall possessed by the proposed approach are due to the working of the intrusion detection system, in which the ranking of features is performed in SIDS to classify the packets based on important features in the attack patterns. The temporal dependencies between the features possessed by the suspicious packets are analyzed to precisely determine the packets. The overall reduction in memory usage, CPU usage and execution time is due to the adoption of edge-based architecture, in which the service migration is carried out based on configuration, resource availability, trust and load of the edge servers. Further, blockchain-based authentication ensures the legitimacy of the entities in the network. From the above discussion, it is clear that the proposed approach performs more effectively than the existing approaches in terms of all the metrics considered, providing security. Table~\ref{TABLE VI: COMPARISON OF PERFORMANCE METRICS} shows the performance metrics comparison of existing and proposed approaches.

\section{CONCLUSION AND FUTURE DIRECTIONS}

The rapid expansion of the IoT within NGWN has introduced significant security challenges. In this paper, we proposed a trust-aware security framework that integrates blockchain authentication, advanced intrusion detection, and dynamic honeypot deception to enhance security in NGWN-enabled IoT environments. Our framework effectively mitigates cybersecurity threats by ensuring robust authentication, reducing false positives in intrusion detection, and dynamically adapting to evolving attack strategies. The proposed system leverages blockchain technology for decentralized and tamper-proof authentication, preventing impersonation and unauthorized access. The integration of an IRF-based SIDS and a DCRNN-based AIDS enables precise and efficient threat detection. Furthermore, a moving target defence strategy dynamically migrates services to trusted edge nodes, mitigating the risk of service compromise. The deployment of high-interaction, on-demand virtual honeypots allows for proactive deception and real-time attack pattern analysis, significantly improving intrusion detection and response mechanisms. Performance evaluation in the NS3 simulation environment demonstrates the superiority of our approach over existing methods. The proposed framework achieves a 25\% improvement in detection accuracy, a 30\% reduction in false negatives, and enhanced resource efficiency compared to conventional security mechanisms. These results confirm the effectiveness of our approach in securing IoT environments while maintaining low computational overhead. Despite its advancements, this research opens avenues for further exploration. Future work will focus on integrating Explainable AI (XAI) techniques to improve transparency and interpretability in intrusion detection decisions. Incorporating XAI, security professionals and system administrators will gain deeper insights into the decision-making process of the detection models, ensuring better trust, usability, and adaptability of the security framework.



\begin{thebibliography}{10}
\providecommand{\url}[1]{#1}
\csname url@samestyle\endcsname
\providecommand{\newblock}{\relax}
\providecommand{\bibinfo}[2]{#2}
\providecommand{\BIBentrySTDinterwordspacing}{\spaceskip=0pt\relax}
\providecommand{\BIBentryALTinterwordstretchfactor}{4}
\providecommand{\BIBentryALTinterwordspacing}{\spaceskip=\fontdimen2\font plus
\BIBentryALTinterwordstretchfactor\fontdimen3\font minus \fontdimen4\font\relax}
\providecommand{\BIBforeignlanguage}[2]{{%
\expandafter\ifx\csname l@#1\endcsname\relax
\typeout{** WARNING: IEEEtran.bst: No hyphenation pattern has been}%
\typeout{** loaded for the language `#1'. Using the pattern for}%
\typeout{** the default language instead.}%
\else
\language=\csname l@#1\endcsname
\fi
#2}}
\providecommand{\BIBdecl}{\relax}
\BIBdecl

\bibitem{zhang2020real}
Y.~Zhang, G.~Chen, H.~Du, X.~Yuan, M.~Kadoch, and M.~Cheriet, ``Real-time remote health monitoring system driven by 5g mec-iot,'' \emph{Electronics}, vol.~9, no.~11, p. 1753, 2020.

\bibitem{wazid2020security}
M.~Wazid, A.~K. Das, S.~Shetty, P.~Gope, and J.~J. Rodrigues, ``Security in 5g-enabled internet of things communication: issues, challenges, and future research roadmap,'' \emph{IEEE Access}, vol.~9, pp. 4466--4489, 2020.

\bibitem{nilkanthsing2020dynamic}
V.~Nilkanthsing, ``Dynamic orchestration of security services at fog nodes for 5g iot,'' in \emph{IEEE International Conference on Communication}, 2020.

\bibitem{aravamudhan2021survey}
P.~Aravamudhan and T.~Kanimozhi, ``A survey on intrusion detection system and prerequisite demands in {IoT} networks,'' in \emph{Journal of Physics: Conference Series}, vol. 1916, no.~1.\hskip 1em plus 0.5em minus 0.4em\relax IOP Publishing, 2021, p. 012179.

\bibitem{chaabouni2019network}
N.~Chaabouni, M.~Mosbah, A.~Zemmari, C.~Sauvignac, and P.~Faruki, ``Network intrusion detection for iot security based on learning techniques,'' \emph{IEEE Communications Surveys \& Tutorials}, vol.~21, no.~3, pp. 2671--2701, 2019.

\bibitem{al2025comprehensive}
Q.~A. Al-Haija and A.~Droos, ``A comprehensive survey on deep learning-based intrusion detection systems in internet of things (iot),'' \emph{Expert Systems}, vol.~42, no.~2, p. e13726, 2025.

\bibitem{hinojosa2024edge}
A.~Hinojosa and N.~E. Majd, ``Edge computing network intrusion detection system in iot using deep learning,'' in \emph{2024 33rd International Conference on Computer Communications and Networks (ICCCN)}.\hskip 1em plus 0.5em minus 0.4em\relax IEEE, 2024, pp. 1--6.

\bibitem{toony2024multi}
A.~A. Toony, F.~Alqahtani, Y.~Alginahi, and W.~Said, ``Multi-block: A novel ml-based intrusion detection framework for sdn-enabled iot networks using new pyramidal structure,'' \emph{Internet of Things}, vol.~26, p. 101231, 2024.

\bibitem{saheed2024modified}
Y.~K. Saheed, O.~H. Abdulganiyu, and T.~A. Tchakoucht, ``Modified genetic algorithm and fine-tuned long short-term memory network for intrusion detection in the internet of things networks with edge capabilities,'' \emph{Applied Soft Computing}, vol. 155, p. 111434, 2024.

\bibitem{maurya2025blockchain}
V.~Maurya, V.~Rishiwal, M.~Yadav, M.~Shiblee, P.~Yadav, U.~Agarwal, and R.~Chaudhry, ``Blockchain-driven security for iot networks: State-of-the-art, challenges and future directions,'' \emph{Peer-to-Peer Networking and Applications}, vol.~18, no.~1, pp. 1--35, 2025.

\bibitem{otoum2024advancing}
Y.~Otoum, P.~Singh, and A.~Nayak, ``Advancing iomt defenses: Deep collaborative learning for robust healthcare security,'' in \emph{GLOBECOM 2024-2024 IEEE Global Communications Conference}.\hskip 1em plus 0.5em minus 0.4em\relax IEEE, 2024, pp. 2966--2971.

\bibitem{dawit2020suitability}
N.~A. Dawit, S.~S. Mathew, and K.~Hayawi, ``Suitability of blockchain for collaborative intrusion detection systems,'' in \emph{2020 12th Annual Undergraduate Research Conference on Applied Computing (URC)}.\hskip 1em plus 0.5em minus 0.4em\relax IEEE, 2020, pp. 1--6.

\bibitem{louati2020deep}
F.~Louati and F.~B. Ktata, ``A deep learning-based multi-agent system for intrusion detection,'' \emph{SN Applied Sciences}, vol.~2, no.~4, p. 675, 2020.

\bibitem{franco2021survey}
J.~Franco, A.~Aris, B.~Canberk, and A.~S. Uluagac, ``A survey of honeypots and honeynets for internet of things, industrial internet of things, and cyber-physical systems,'' \emph{IEEE Communications Surveys \& Tutorials}, vol.~23, no.~4, pp. 2351--2383, 2021.

\bibitem{khan2020reputation}
Z.~A. Khan and U.~Abbasi, ``Reputation management using honeypots for intrusion detection in the internet of things,'' \emph{Electronics}, vol.~9, no.~3, p. 415, 2020.

\bibitem{zheng2025predictive}
T.~Zheng, Y.~Du, K.~Hua, X.~Wu, S.~Yuan, X.~Wang, Q.~Chen, and J.~Tan, ``Predictive analytics for cyber-attack timing in power internet of things: A flipit game-theoretic approach,'' \emph{Internet of Things}, p. 101522, 2025.

\bibitem{otoum2024enhancing}
Y.~Otoum, C.~Hu, E.~H. Said, and A.~Nayak, ``Enhancing heart disease prediction with federated learning and blockchain integration,'' \emph{Future Internet}, vol.~16, no.~10, p. 372, 2024.

\bibitem{parsamehr2019novel}
R.~Parsamehr, A.~Esfahani, G.~Mantas, A.~Radwan, S.~Mumtaz, J.~Rodriguez, and J.-F. Mart{\'\i}nez-Ortega, ``A novel intrusion detection and prevention scheme for network coding-enabled mobile small cells,'' \emph{IEEE Transactions on Computational Social Systems}, vol.~6, no.~6, pp. 1467--1477, 2019.

\bibitem{liang2020intrusion}
C.~Liang, B.~Shanmugam, S.~Azam, A.~Karim, A.~Islam, M.~Zamani, S.~Kavianpour, and N.~B. Idris, ``Intrusion detection system for the internet of things based on blockchain and multi-agent systems,'' \emph{Electronics}, vol.~9, no.~7, p. 1120, 2020.

\bibitem{mamolar2019autonomic}
A.~S. Mamolar, P.~Salv{\'a}-Garc{\'\i}a, E.~Chirivella-Perez, Z.~Pervez, J.~M.~A. Calero, and Q.~Wang, ``Autonomic protection of multi-tenant 5g mobile networks against udp flooding ddos attacks,'' \emph{Journal of Network and Computer Applications}, vol. 145, p. 102416, 2019.

\bibitem{rajkumar2025multi}
M.~Rajkumar, J.~Karthika \emph{et~al.}, ``Multi-view consistent generative adversarial network for enhancing intrusion detection with prevention systems in mobile ad hoc networks against security attacks,'' \emph{Computers \& Security}, vol. 150, p. 104242, 2025.

\bibitem{al2020real}
Y.~Al-Hadhrami and F.~K. Hussain, ``Real time dataset generation framework for intrusion detection systems in iot,'' \emph{Future Generation Computer Systems}, vol. 108, pp. 414--423, 2020.

\bibitem{hatzivasilis2019wardog}
G.~Hatzivasilis, O.~Soultatos, P.~Chatziadam, K.~Fysarakis, I.~Askoxylakis, S.~Ioannidis, G.~Alexandris, V.~Katos, and G.~Spanoudakis, ``Wardog: Awareness detection watchdog for botnet infection on the host device,'' \emph{IEEE Transactions on Sustainable Computing}, vol.~6, no.~1, pp. 4--18, 2019.

\bibitem{wang2018secure}
W.~Wang, P.~Xu, and L.~T. Yang, ``Secure data collection, storage and access in cloud-assisted iot,'' \emph{IEEE cloud computing}, vol.~5, no.~4, pp. 77--88, 2018.

\bibitem{maimo2018self}
L.~F. Maim{\'o}, {\'A}.~L.~P. G{\'o}mez, F.~J.~G. Clemente, M.~G. P{\'e}rez, and G.~M. P{\'e}rez, ``A self-adaptive deep learning-based system for anomaly detection in 5g networks,'' \emph{Ieee Access}, vol.~6, pp. 7700--7712, 2018.

\bibitem{lam2020machine}
J.~Lam and R.~Abbas, ``Machine learning based anomaly detection for 5g networks,'' \emph{arXiv preprint arXiv:2003.03474}, 2020.

\bibitem{yang2019design}
A.~Yang, Y.~Zhuansun, C.~Liu, J.~Li, and C.~Zhang, ``Design of intrusion detection system for internet of things based on improved bp neural network,'' \emph{Ieee Access}, vol.~7, pp. 106\,043--106\,052, 2019.

\bibitem{mondal2021enhanced}
A.~Mondal and R.~T. Goswami, ``Enhanced honeypot cryptographic scheme and privacy preservation for an effective prediction in cloud security,'' \emph{Microprocessors and Microsystems}, vol.~81, p. 103719, 2021.

\bibitem{eskandari2020passban}
M.~Eskandari, Z.~H. Janjua, M.~Vecchio, and F.~Antonelli, ``Passban ids: An intelligent anomaly-based intrusion detection system for iot edge devices,'' \emph{IEEE Internet of Things Journal}, vol.~7, no.~8, pp. 6882--6897, 2020.

\bibitem{li2020deepfed}
B.~Li, Y.~Wu, J.~Song, R.~Lu, T.~Li, and L.~Zhao, ``Deepfed: Federated deep learning for intrusion detection in industrial cyber--physical systems,'' \emph{IEEE Transactions on Industrial Informatics}, vol.~17, no.~8, pp. 5615--5624, 2020.

\bibitem{mahdi2024secure}
M.~A. Mahdi, ``Secure and efficient iot networks: An ai and ml-based intrusion detection system,'' in \emph{2024 3rd International Conference on Artificial Intelligence For Internet of Things (AIIoT)}.\hskip 1em plus 0.5em minus 0.4em\relax IEEE, 2024, pp. 1--6.

\bibitem{mallidi2025advancements}
S.~K.~R. Mallidi and R.~R. Ramisetty, ``Advancements in training and deployment strategies for ai-based intrusion detection systems in iot: a systematic literature review,'' \emph{Discover Internet of Things}, vol.~5, no.~1, p.~8, 2025.

\bibitem{al2020analysis}
H.~Al-Mohannadi, I.~Awan, and J.~Al~Hamar, ``Analysis of adversary activities using cloud-based web services to enhance cyber threat intelligence,'' \emph{Service Oriented Computing and Applications}, vol.~14, no.~3, pp. 175--187, 2020.

\bibitem{lee2020phantomfs}
J.~Lee, J.~Choi, G.~Lee, S.-W. Shim, and T.~Kim, ``Phantomfs: File-based deception technology for thwarting malicious users,'' \emph{IEEE Access}, vol.~8, pp. 32\,203--32\,214, 2020.

\bibitem{li2020anti}
B.~Li, Y.~Xiao, Y.~Shi, Q.~Kong, Y.~Wu, and H.~Bao, ``Anti-honeypot enabled optimal attack strategy for industrial cyber-physical systems,'' \emph{IEEE Open Journal of the Computer Society}, vol.~1, pp. 250--261, 2020.

\bibitem{dara2024intelligent}
N.~Dara, P.~Shankar, P.~V. Arvind, and V.~Singh, ``Intelligent insight into iot threats: Leveraging advanced analytics with honeypots for anomaly detection,'' in \emph{2024 IEEE 9th International Conference for Convergence in Technology (I2CT)}.\hskip 1em plus 0.5em minus 0.4em\relax IEEE, 2024, pp. 1--6.

\bibitem{ntizikira2024honey}
E.~Ntizikira, L.~Wang, J.~Chen, and K.~Saleem, ``Honey-block: Edge assisted ensemble learning model for intrusion detection and prevention using defense mechanism in iot,'' \emph{Computer Communications}, vol. 214, pp. 1--17, 2024.

\bibitem{jangirala2019designing}
S.~Jangirala, A.~K. Das, and A.~V. Vasilakos, ``Designing secure lightweight blockchain-enabled rfid-based authentication protocol for supply chains in 5g mobile edge computing environment,'' \emph{IEEE Transactions on Industrial Informatics}, vol.~16, no.~11, pp. 7081--7093, 2019.

\bibitem{wahab2019resource}
O.~A. Wahab, J.~Bentahar, H.~Otrok, and A.~Mourad, ``Resource-aware detection and defense system against multi-type attacks in the cloud: Repeated bayesian stackelberg game,'' \emph{IEEE Transactions on Dependable and Secure Computing}, vol.~18, no.~2, pp. 605--622, 2019.

\bibitem{gill2020gtm}
K.~S. Gill, S.~Saxena, and A.~Sharma, ``Gtm-csec: Game theoretic model for cloud security based on ids and honeypot,'' \emph{Computers \& Security}, vol.~92, p. 101732, 2020.

\bibitem{chakkaravarthy2020design}
S.~S. Chakkaravarthy, D.~Sangeetha, M.~V. Cruz, V.~Vaidehi, and B.~Raman, ``Design of intrusion detection honeypot using social leopard algorithm to detect iot ransomware attacks,'' \emph{IEEE Access}, vol.~8, pp. 169\,944--169\,956, 2020.

\bibitem{zhang2019iot}
W.~Zhang, B.~Zhang, Y.~Zhou, H.~He, and Z.~Ding, ``An iot honeynet based on multiport honeypots for capturing iot attacks,'' \emph{IEEE Internet of Things Journal}, vol.~7, no.~5, pp. 3991--3999, 2019.

\end{thebibliography}

\end{document}